\documentclass{ieeeaccess}
\usepackage{cite}
\usepackage{amsmath,amssymb,amsfonts}
\usepackage{algorithmic}
\usepackage{graphicx}
\usepackage{textcomp}
\usepackage{float}
\usepackage{multirow}
\usepackage{multicol}
\usepackage{booktabs}
\usepackage{multirow}
\usepackage{longtable}
\usepackage{subcaption}

\usepackage{caption,setspace}
\def\BibTeX{{\rm B\kern-.05em{\sc i\kern-.025em b}\kern-.08em
    T\kern-.1667em\lower.7ex\hbox{E}\kern-.125emX}}
\begin{document}
\history{}
\doi{}

\title{Attitude Control and Low Cost Design of UAV Bicopter}
\author{\uppercase{Fahmizal}\authorrefmark{1,$\circ$,$\lozenge$}, 
\uppercase{Hanung Adi Nugroho}\authorrefmark{2,$\circ$}, 
\uppercase{Adha Imam Cahyadi}\authorrefmark{3,$\circ$},
and \uppercase{Igi Ardiyanto}\authorrefmark{4,$\circ$}}

\address[$\circ$]{Department of Electrical and Information Engineering, Engineering Faculty, Universitas Gadjah Mada, 55281, Indonesia}
\address[$\lozenge$]{Department of Electrical Engineering and Informatics, Vocational College, Universitas Gadjah Mada, 55281, Indonesia}

\markboth
{Fahmizal \headeretal: Attitude Control and Low Cost Design of UAV Bicopter}
{Fahmizal \headeretal: Attitude Control and Low Cost Design of UAV Bicopter}

\corresp{Email: $^{1}$fahmizal@ugm.ac.id, $^{2}$adinugroho@ugm.ac.id, $^{3}$adha.imam@ugm.ac.id, $^{4}$igi@ugm.ac.id}

\begin{abstract}
This paper present a control system for the attitude and low cost design of a Bicopter. The control system uses a PID controller that receives feedback from an IMU to calculate control inputs that adjust the Bicopter's attitude (roll, pitch and yaw angles) which is resistant to disturbances (wind noise) on a test bed. The control system is implemented on a hardware platform consisting of a Bicopter, an IMU sensor, and a microcontroller with low cost design. In mechanical design, the Bicopter is designed to more closely resemble the letter "V" so that the distribution of the centre of mass (CoM) of the Bicopter can be such that the servomotor torque reaction is parallel to the axis of rotation of the Bicopter during the movement of the pitch angle attitude. In electronic design, the Bicopter was developed using the ATmega328P microcontroller.
\end{abstract}


\titlepgskip=-21pt

\maketitle

\section{Introduction}
\label{sec:introduction}
\normalfont Unmanned aerial vehicles (UAV) Bicopters are becoming increasingly popular due to their ability to perform a wide range of tasks, such as aerial photography, monitoring, and surveying \cite{kawasaki2015dual}. The unique design of Bicopters, which combines the features of both fixed-wing and rotary-wing aircraft, makes them well suited to many applications \cite{saeed2018survey,ke2018design}. However, controlling the attitude of a Bicopter can be challenging due to the complex and nonlinear dynamics involved. 

The control of attitude is a critical issue in the design and operation of UAVs. In particular, for a UAV Bicopter, which has a hybrid design combining the advantages of a helicopter and a fixed-wing aircraft, controlling the attitude is essential for stable flight and manoeuvrability. Conventional control approaches, such as proportional-integral-derivative (PID) controllers, have been widely used for Bicopter attitude control \cite{qin2020gemini,albayrak2019design,zhang2016modeling,hrevcko2015bicopter,panigrahi2021design}. These controllers are easy to implement and have been shown to be effective in many cases. 

Advanced model-based controllers, such as linear quadratic regulators (LQR), have been proposed as an alternative to PID controllers \cite{araar2014full,cohen2020finite,heng2015trajectory,okyere2019lqr,saraf2020comparative}. These controllers use a mathematical model of the Bicopter to predict its behavior and adjust the control inputs accordingly. While LQR controllers can be more effective than PID controllers in some cases, they are also more complex and require more computational resources.

Youming Qin et al. \cite{qin2020gemini} detailed a Bicopter concept they called Gemini in their research. In this research, they focus on how the Bicopter can be put to use in enclosed environments. Starting with the selection of the optimal propeller and continuing through aerodynamic analysis, system design, optimisation, and implementation of the controls, this study details the full process of creating the Gemini platform. Cascaded PID controllers are used in practise on Gemini's attitude controller. In this research use high cost flight-controller.

In 2022, Bicopter's research on mechanical design without a servomotor has been carried out by Youming Qin et al. \cite{qin2022gemini}. In this study, replacing the servomotor by using a cyclic blade pitch control but requires a magnetic encoder sensor on the cyclic blade system so that it becomes a high cost flight controller.

This paper has the following contributions: 1). developing and implementing a PID controller design to maintain a stable attitude of the UAV Bicopter, which is resistant to disturbances (wind noise) on a test bed. 2). designs and manufactures mechanical (tilt mechanism) and electronic (flight controller) UAV Bicopter with the principle of low cost.

This paper's remaining sections are organized as follows: Section II covers the methodology of mechanical design, electronics design, attitude sensor, and attitude modelling. Section III describes the design of the attitude control using the PID controller. Experimental results are presented in Section IV to demonstrate the value and effectiveness of the proposed methodologies. Section V concludes the paper.

\newpage
\section{Materials and Methods}
\subsection{Mechanical Design}

Bicopters are a type of UAV that have two rotors that are fixed in a parallel configuration, which allows them to perform VTOL and hover like a helicopter, as well as to fly forward like a fixed-wing aircraft. Designing the mechanics of a Bicopter involves several considerations, including the size and weight of the vehicle, the choice of materials, the design of the rotors, and the placement of the motors. The size and weight of the Bicopter will determine the amount of lift required and the amount of power needed to achieve flight. The size and weight will also determine the maximum payload capacity of the Bicopter. The rotors are a critical component of the Bicopter, as they provide lift and control the vehicle's attitude.

\begin{figure}[h]
	\centering
	\includegraphics[scale=0.3]{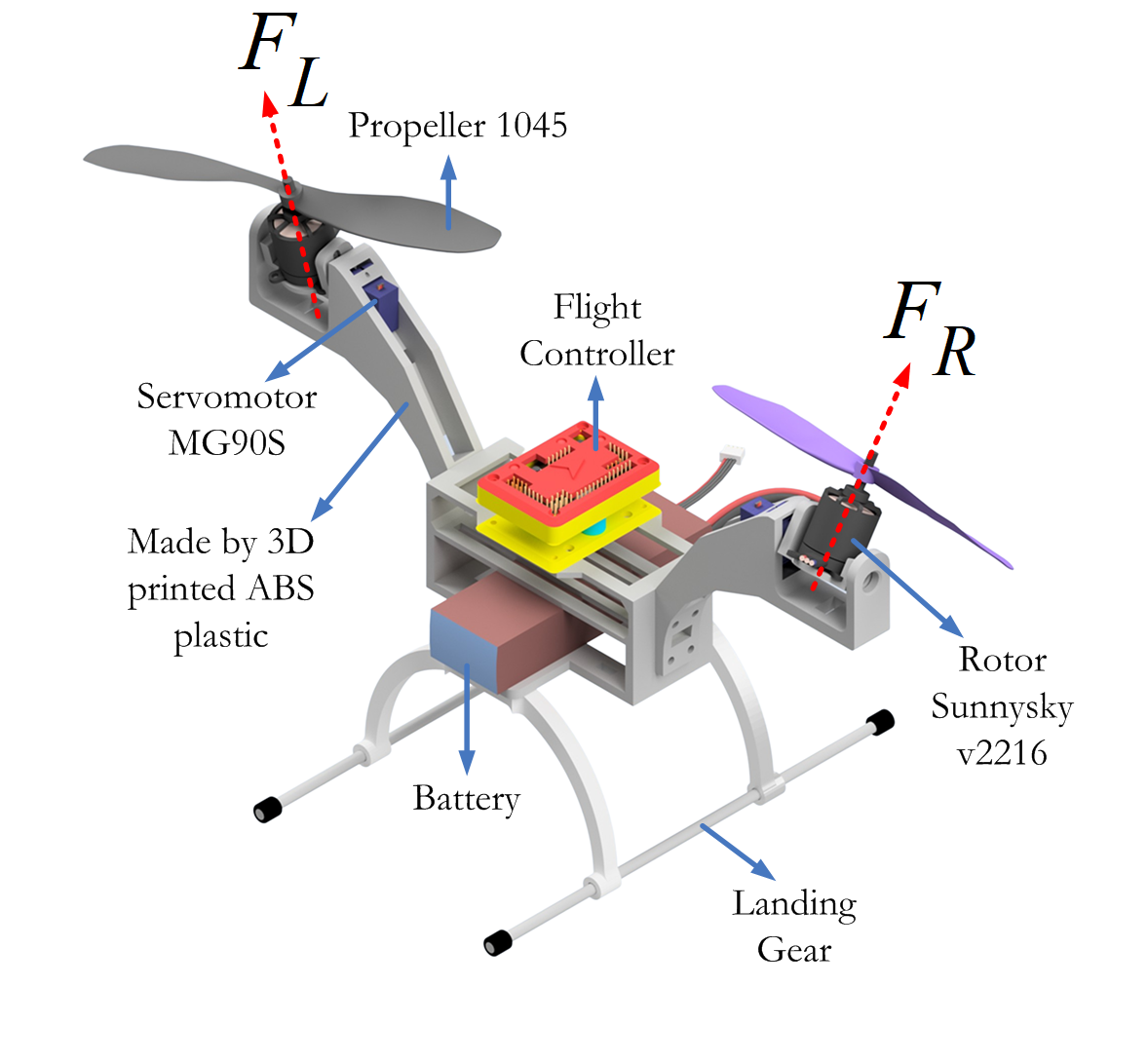}
	\caption{Mechanical design of Bicopter.}
	\label{FullDesainMekanis}
\end{figure}

The Bicopter is constructed with two driving rotors and two servomotors for tilting the two rotors in opposite directions. Figure \ref{FullDesainMekanis} shows the right rotor thrust ($F_{R}$) and left rotor thrust ($F_{L}$) created by the rotor, propeller, and their components in the $x$ and $z$ axes. By altering the magnitude of the rotor thrust $F_{R}$ and $F_{L}$, the rolling movement may be adjusted. This paper develops the mechanical design of Bicopter using Autodesk Inventor Student version and printed by UP2 3D printer. The mechanical design of the UAV Bicopter consists of an arm, rotor holder and body. As shown in Fig. \ref{FullDesainMekanis}, the Bicopter is designed to more closely resemble the letter "V" so that the distribution of the centre of mass (CoM) of the Bicopter can be such that the servomotor torque reaction is parallel to the axis of rotation of the Bicopter during the movement of the pitch angle attitude.

\begin{figure}[h]
	\centering
	\includegraphics[scale=0.38]{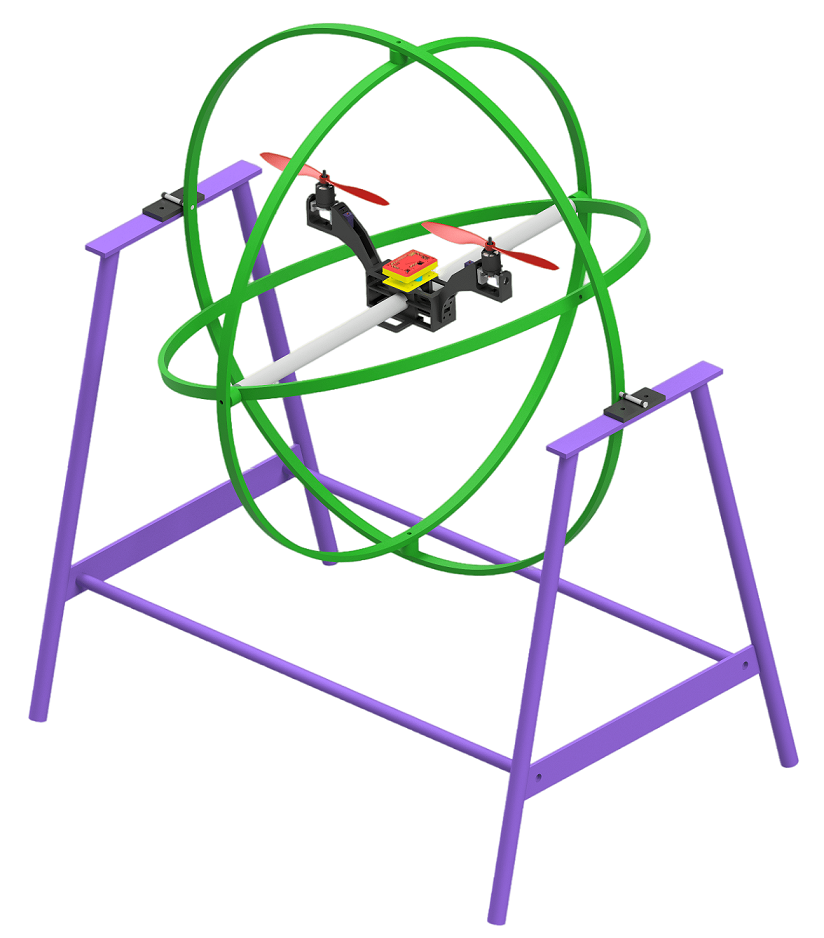}
	\caption{Test bed rig for attitude testing of Bicopter.}
	\label{fig:testbed}
\end{figure}

The test bed rig is a device used as an evaluation platform for the stability of the rotational motion of roll, pitch, and yaw on a Bicopter. Without needing to fly the Bicopter, the stability of its attitude can be verified with this test bed rig. The Bicopter's attitude when disturbed can also be observed with this test bed rig. Figure \ref{fig:testbed} illustrates the design of the test bed rig that will be used for Bicopter attitude testing.

\subsection{Electronics Design}
\begin{figure}[h]
	\centering
	\includegraphics[scale=0.26]{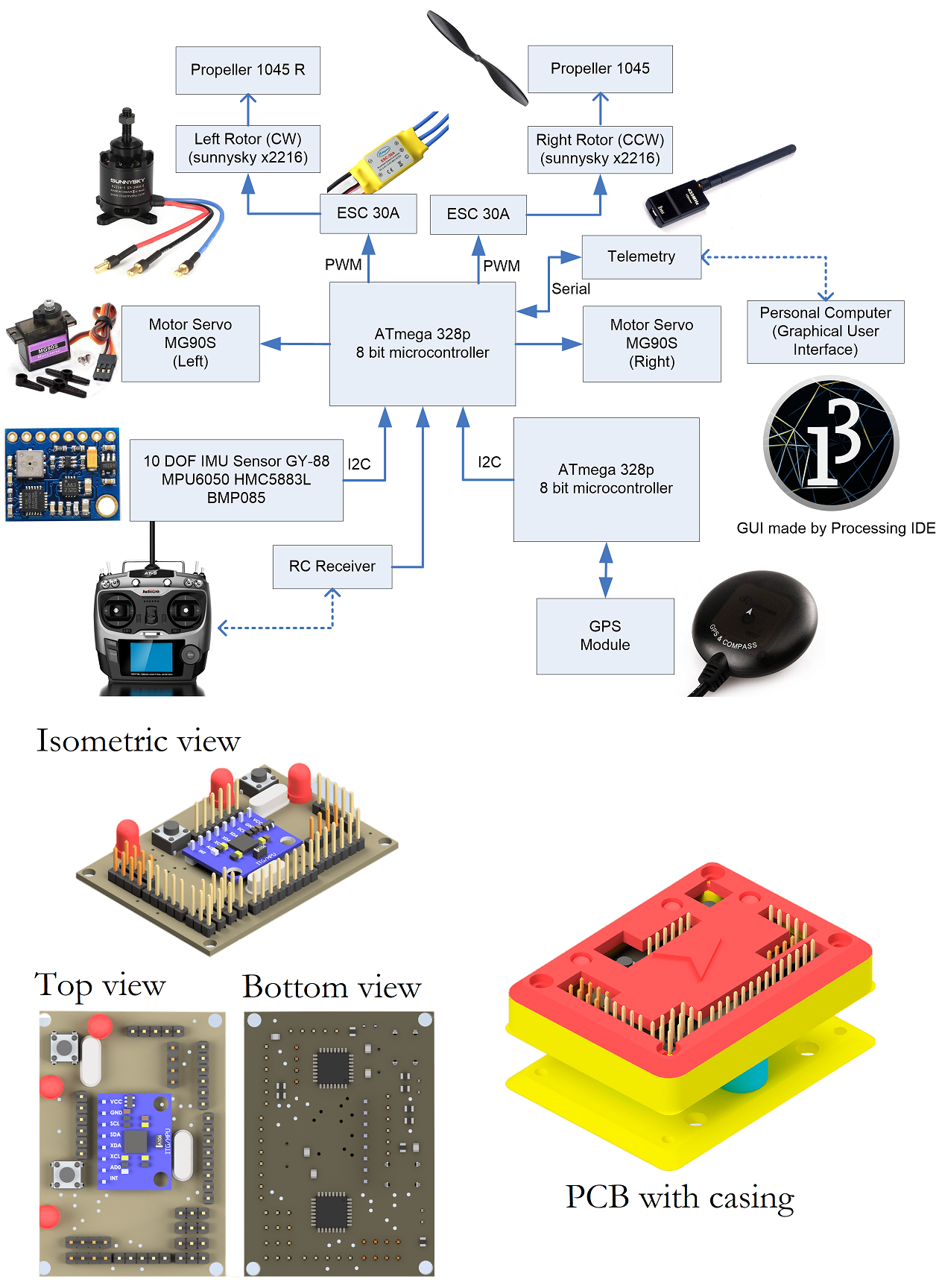}
	\caption{Electronic design of Bicopter.}
	\label{desainelektronis}
\end{figure}

\begin{figure}[h]
	\centering
	\includegraphics[scale=0.33]{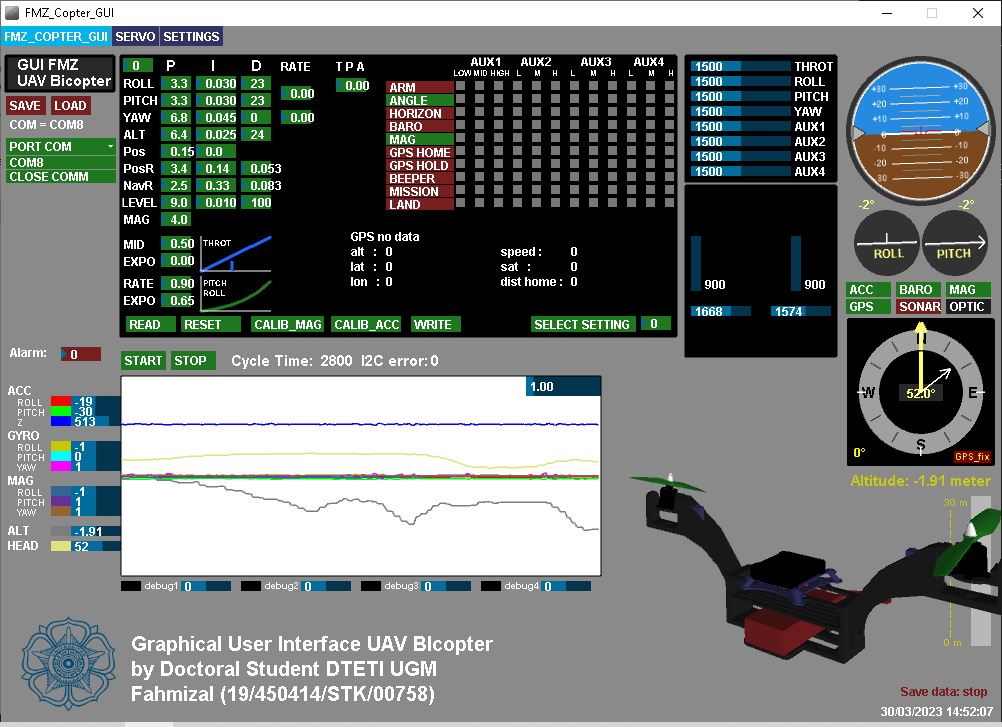}
	\caption{Graphical user interface (GUI) of Bicopter.}
	\label{GUI}
\end{figure}

ATmega328P serves as the primary microcontroller in the Bicopter electronics system. There are two MG90S servomotors and two left and right rotors use Sunnysky x2216 800 KV rotors, and one IMU sensor MPU6050.  Figure \ref{desainelektronis} represents the results of the printed circuit board (PCB) design for the Bicopter electronic system. This PCB is also known as the Bicopter's flight controller. The electronic system is also coupled with a personal computer (PC) via serial communication with the graphical user interface (GUI) to automatically show sensor reading conditions in real time as presented in Fig. \ref{GUI} .

\subsection{Attitude Sensor}

The \textit{motion processing unit} (MPU) 6050 is a type of inertial measurement unit (IMU) that is commonly used in small UAVs and other electronic devices that require accurate motion sensing. It is a small and affordable sensor that combines both accelerometer and gyroscope functionality. The accelerometer measures acceleration along the $x$, $y$, and $z$ axes and is used to determine the orientation of the device. It senses both static and dynamic acceleration due to gravity and motion respectively. It is able to measure acceleration in the range of $\pm 2g$, $\pm 4g$, $\pm 8g$, or $\pm 16g$.

In this paper, the IMU MPU 6050 is used in the Bicopter orientation sensor system. This sensor's configuration is as follows: it has six degrees of freedom and is made up of two types of sensors, namely accelerometers and gyroscopes with data transmission protocols based on inter-integrated circuits (I2C) \cite{fedorov2015using}. The IMU MPU 6050 sensor produces an angle on the x-axis called the $phi$ ($\phi$) or roll angle, a pitch angle on the y-axis called the $theta$ ($\theta$), and a yaw angle on the z-axis called the $psi$ ($\psi $).

\begin{figure}[H]
	\centering
	\includegraphics[scale=0.18]{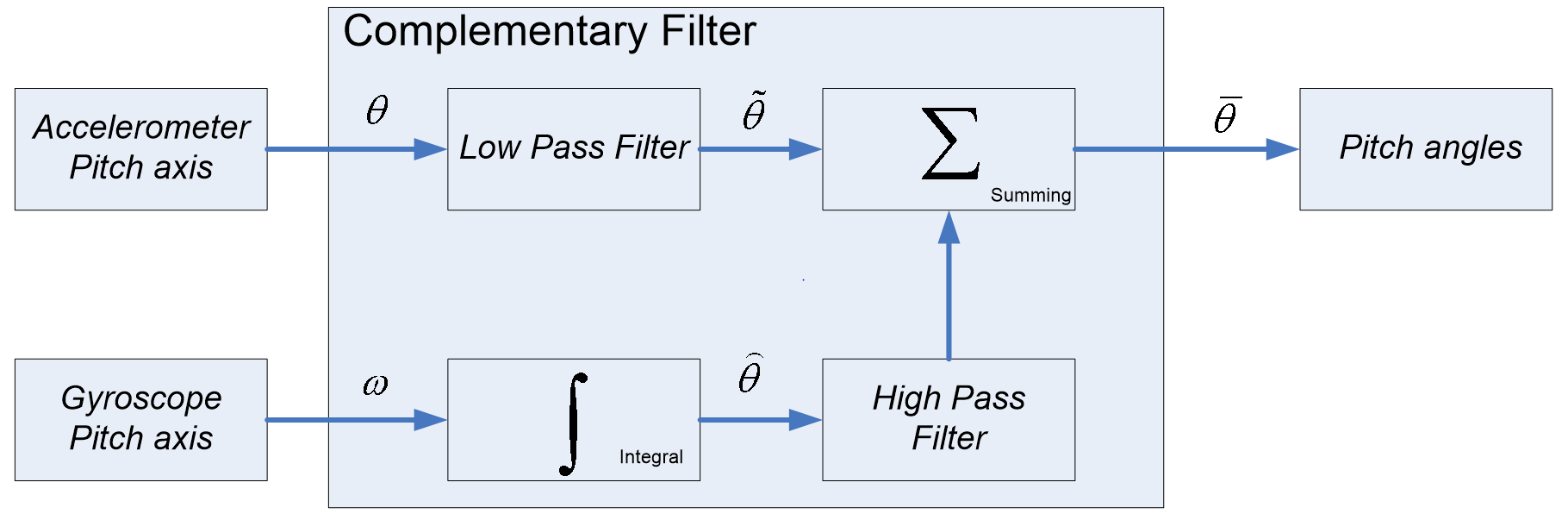}
	\caption{Complementary filter block diagram.}
	\label{diagramCF}
\end{figure}

Noise is an important distraction that must be considered in a measuring process. Besides that, noise can also interfere with the process in a closed-loop control system. Because of that, filtering techniques are needed to separate the actual signal from a set of noises. This paper uses a complementary filter (CF) to remove noise \cite{berkane2017design,ngo2017experimental}. The configuration of the CF for the pitch angle is presented in Fig. \ref{diagramCF}. The characteristics of the raw accelerometer sensor data are filtered in the low-frequency region, and the gyroscope data are in the high-frequency region \cite{marantos2015uav}. Therefore, to compensate for the sensor data, apply a low-pass filter (LPF) and a high-pass filter (HPF). Theoretically, an LPF is shown in Eq. (\ref{eq2w17}).

\begin{equation}
	V_{in}(t)-V_{out}(t)=RC\frac{dV_{out}}{dt}
	\label{eq2w17}
\end{equation}

Equation \ref{eq2w17} can be discretized into Eq. (\ref{eq2w18}). Furthermore, for simplicity, it is assumed that the input and output are sampled at the same time interval, namely $\Delta _{T}$. Input $V_{in}$ is defined by $x_{i}$ and output $V_{out}$ is defined by $y_{i}$.

\begin{align}
	&x_{i}-y_{i}=RC\frac{y_{i}-y_{i-1}}{\Delta _{T}} \nonumber \\
	&y_{i}=x_{i}\left ( \frac{\Delta _{T}}{RC+\Delta _{T}} \right )+y_{i-1}\left ( \frac{RC}{RC+\Delta _{T}}\right )\nonumber \\
	&y_{i}=\alpha x_{i}+\left ( 1-\alpha  \right )y_{i-1}
	\label{eq2w18}
\end{align}

Where $\alpha = \frac{\Delta _{T}}{RC+\Delta _{T}}$, $RC=\frac{1}{2 \pi f_{c}}$ , $f_{c}$ show as frequency \textit{cut-off}, $\Delta _{T}$ is the sampling period and the smoothing factor lies between $0\leq \alpha \leq 1$, then HPF is defined as in Eq. (\ref{eq2w19}).

\begin{align}
	&y_{i}=RC\left ( \frac{x_{i}-x_{i-1}}{\Delta _{T}} -\frac{y_{i}-y_{i-1}}{\Delta _{T}}\right ) \nonumber \\
	&y_{i}=\left ( \frac{\Delta_{T}}{RC+\Delta_{T}} \right )y_{i-1}+\left ( \frac{RC}{RC+\Delta_{T}} \right )\left ( x_{i}-x_{i-1} \right )\nonumber \\
	&y_{i}=\alpha y_{i-1}+\alpha \left ( x_{i}-x_{i-1} \right )
	\label{eq2w19}
\end{align}

In addition to implementing the IMU MPU 6050 using CF, this paper also uses a Quaternion-based sensor fusion processing technique. Quaternions were introduced to improve computational efficiency and to avoid gimbal lock and singularity \cite{fourati2012complementary, hemingway2018perspectives} problems in the Euler angle representation \cite{brigante2011towards,hua2013implementation,oh2014attitude}. Quaternions have four dimensions, one real dimension (scalar element) and three imaginary dimensions ( vector). Each of these imaginary dimensions has a unit value of the square root of -1, but the square roots of the distinct -1 are perpendicular to one another; we know $i$, $j$ and $k$. So, a Quaternion can be represented as $q=w+ix+jy+kz$. A Quaternion can be converted into a 3D space like Euler and the yaw-pitch-roll (YPR) representation \cite{katsuki2015rotation, parwana2017quaternions}. This makes it easier to imagine and describe rotations along the $x$, $y$, and $z$ axes. First, by extracting the gravity $g=\begin{bmatrix}
	g_{x} &g_{y}  &g_{z} 
\end{bmatrix}$ from the Quaternion defined by Eq. (\ref{eq2w20}).

\begin{equation}
	\begin{bmatrix}
		g_{x}\\ 
		g_{y}\\ 
		g_{z}
	\end{bmatrix}=\begin{bmatrix}
		2\left ( q_{x}q_{z}-q_{w}q_{y} \right )\\ 
		2\left ( q_{w}q_{x}-q_{y}q_{z} \right )\\ 
		q_{w}q_{w}-q_{x}q_{x}+q_{z}q_{z}
	\end{bmatrix}
	\label{eq2w20}
\end{equation}

Then YPR and Euler can be obtained by conversion in Eq. (\ref{eq2w21}) and Eq. (\ref{eq2w22}).

\begin{equation}
	\begin{bmatrix}
		yaw\\ 
		pitch\\ 
		roll
	\end{bmatrix}=\begin{bmatrix}
		\arctan 2\left ( \frac{2q_{x}q_{y}-2q_{w}q_{z}}{2q_{w}q_{w}+2q_{x}q_{x}^{-1}} \right )\\ 
		\arctan \left ( \frac{g_{w}}{\sqrt{g_{x}g_{x}+g_{y}g_{y}}} \right )\\ 
		\arctan\left ( \frac{g_{x}}{\sqrt{g_{w}g_{w}+g_{y}g_{y}}} \right )
	\end{bmatrix}
	\label{eq2w21}
\end{equation}

\begin{equation}
	\begin{bmatrix}
		\psi \\ 
		\theta \\ 
		\phi 
	\end{bmatrix}=\begin{bmatrix}
		\arctan 2\left ( \frac{2q_{x}q_{y}-2q_{w}q_{z}}{2q_{w}q_{w}+2q_{x}q_{x}^{-1}} \right )\\ 
		-\arcsin \left ( 2q_{x}q_{z}+2q_{w}q_{y} \right )\\ 
		\arctan 2\left ( \frac{2q_{y}q_{z}-2q_{w}q_{x}}{2q_{w}q_{w}+2q_{z}q_{z}^{-1}} \right )
	\end{bmatrix}
	\label{eq2w22}
\end{equation}

\subsection{Attitude Modelling}

\begin{figure}[h]
	\centering
	\includegraphics[scale=0.25]{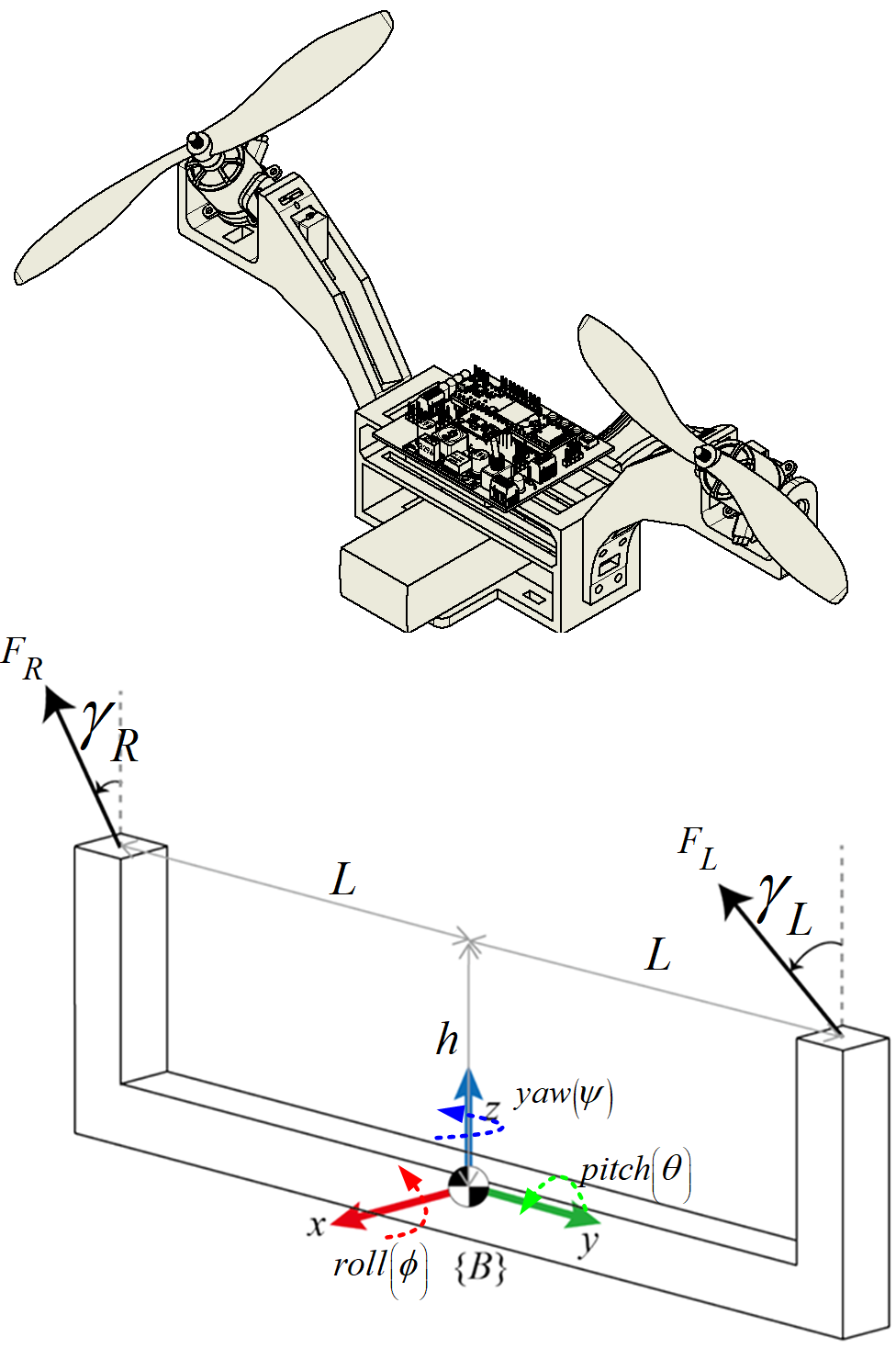}
	\caption{Bicopter reference frames.}
	\label{fig:arahforce}
\end{figure}

The right rotor thrust ($F_{R}$) and left rotor thrust ($F_{L}$) are generated by the propeller and rotor and their components in the $x$ and $z$ directions shown in Fig. \ref{fig:arahforce}. With the parameters described in Table \ref{tab:parameterBicopter}. Using Newton's second law, the equations of forces in the $x$, $y$ and $z$ directions are defined as given in Eq. (\ref{eq3w34}) - (\ref{eq3w36}).

\begin{equation}
	\sum F_{x}=F_{R}\sin \gamma  _{R}+F_{L}\sin \gamma  _{L}
	\label{eq3w34}
\end{equation}
\begin{equation}
	\sum F_{y}=0
	\label{eq3w35}
\end{equation}
\begin{equation}
	\sum F_{z}=F_{R}\cos \gamma  _{R}+F_{L}\cos \gamma  _{L}
	\label{eq3w36}
\end{equation}

\begin{table}[H]
	\centering
	\caption{Bicopter dynamic model parameters.}
	\label{tab:parameterBicopter}
	\resizebox{\columnwidth}{!}{%
		\begin{tabular}{p{0.4\linewidth}ccl}
			\hline
			Parameter                                               & Symbols & Value  & \multicolumn{1}{c}{Unit} \\ \hline
			Mass of the UAV Bicopter                                & $m$       & 0.725   & $kg$                       \\
			Gravitational acceleration                              & $g$       & 9.81   & $m.s^{-2}$                     \\
			Vertical distance between CoG and center   of the rotor & $h$       & 0.042  & $m$                        \\
			Horizontal distance CoG and rotor   center              & $L$       & 0.225  & $m$                        \\
			Thrust coefficient                                      & $C_T$      & 0.1222 &   -                       \\
			The Moment of Inertia along x axis                      & $I_{xx}$     & $0.116 \times10^{-3}$  & $kg.m^{2}$                    \\
			The Moment of Inertia along y axis                      & $I_{yy}$     & $0.0408 \times10^{-3}$ & $kg.m^{2}$                    \\
			The Moment of Inertia along z axis                      & $I_{zz}$     & $0.105 \times10^{-3}$  & $kg.m^{2}$                    \\ \hline
		\end{tabular}%
	}
\end{table}

The total lift (thrust) and moment of force from the Bicopter can be obtained from the input $u$ which is written in Eq. (\ref{eq3w37}) - (\ref{eq3w41}). Where, $C_{T}$ is the thrust coefficient of the propeller. $\Omega _{R}$ and $\Omega _{L}$ are the rotational speeds of the right and left rotors, $\gamma  _{R}$ and $\gamma  _{L}$ are the tilt angles of the right and left rotors.

\begin{equation}
	u=\left [ \begin{matrix}
		u_{1} & u_{2} & u_{3} & u_{4}
	\end{matrix} \right ]^{T}
	\label{eq3w37}
\end{equation}
\begin{equation}
	u_{1}=C_{T}\left ( \Omega _{R}^{2}\cos \gamma  _{R} +\Omega _{L}^{2}\cos\gamma  _{L}\right )
	\label{eq3w38}
\end{equation}
\begin{equation}
	u_{2}=C_{T}\left ( \Omega _{R}^{2}\cos \gamma  _{R} -\Omega _{L}^{2}\cos\gamma  _{L}\right )
	\label{eq3w39}
\end{equation}
\begin{equation}
	u_{3}=C_{T}\left ( \Omega _{R}^{2}\sin \gamma  _{R} +\Omega _{L}^{2}\sin\gamma  _{L}\right )
	\label{eq3w40}
\end{equation}
\begin{equation}
	u_{4}=C_{T}\left ( \Omega _{R}^{2}\sin \gamma  _{R} -\Omega _{L}^{2}\sin\gamma  _{L}\right )
	\label{eq3w41}
\end{equation}

Bicopter dynamic movement can be divided into two subsystems, namely, the rotation subsystem ($roll$, $pitch$ and $yaw$) as the inner loop and the translation subsystem ($x$, $y$ position and $z$ (altitude) position) as the outer loop. Based on the dynamic solution of the model using Newton-Euler \cite{albayrak2019design,zhang2016modeling, abedini2021robust}, we get the equation of translational motion in Eq. (\ref{eq3w42}) and Eq. (\ref{eq3w43}) for rotational motion, where $s=\sin$ and $c=\cos$.

\begin{align}
	&\ddot{x}=-\frac{1}{m}\left ( s\phi s\psi +c\phi s\theta c\psi  \right )u_{1}-\frac{c\theta c\psi }{m}u_{3} \nonumber \\
	&\ddot{y}=-\frac{1}{m}\left ( -s\phi c\psi +c\phi s\theta s\psi  \right )u_{1}+\frac{c\theta s\psi }{m}u_{3} \nonumber \\
	&\ddot{z}=g-\frac{1}{m}\left ( c\phi c\theta  \right )u_{1}-\frac{s\theta }{m}u_{3}
	\label{eq3w42}
\end{align}

\begin{align}
	&\ddot{\phi }=\frac{L}{I_{xx}}u_{2} \nonumber \\
	&\ddot{\theta }=\frac{h}{I_{yy}}u_{3} \nonumber \\
	&\ddot{\psi }=\frac{L}{I_{zz}}u_{4}
	\label{eq3w43}
\end{align}

In this paper, the design of attitude control is the main point, based on the illustration in Fig. \ref{fig:BicopterRoll}, the roll angle movement condition occurs when there is a difference in the lift force caused by the rotation of the right rotor and the left rotor, and the condition of the right and left servomotors is zero ($\cos(0)=1$) so that the rolling case can be solved in Eq. (\ref{eq3w45}). And in the case of pitching and yawing presented in Eq. (\ref{eq3w46}) and Eq. (\ref{eq3w47}). With parameters described in Table \ref{tab:parameterBicopter}.

\begin{equation}
	\ddot{\phi }=\frac{L}{I_{xx}}C_{T}\left ( \Omega _{R}^{2}-\Omega _{L}^{2} \right )=\frac{L}{I_{xx}}C_{T}\left ( F_{R}-F_{L} \right )
	\label{eq3w45}
\end{equation}
\begin{equation}
	\ddot{\theta }=\frac{h}{I_{yy}}C_{T}\left ( \Omega _{R}^{2}\sin \lambda _{R}+ \Omega _{L}^{2}\sin \lambda _{L}\right )
	\label{eq3w46}
\end{equation}
\begin{equation}
	\ddot{\psi }=\frac{L}{I_{zz}}C_{T}\left ( \Omega _{R}^{2}\sin \lambda _{R}-\Omega _{L}^{2}\sin \lambda _{L}\right )
	\label{eq3w47}
\end{equation}

\section{Attitude Control Design}

\begin{figure*}
	\centering
	\includegraphics[scale=0.45]{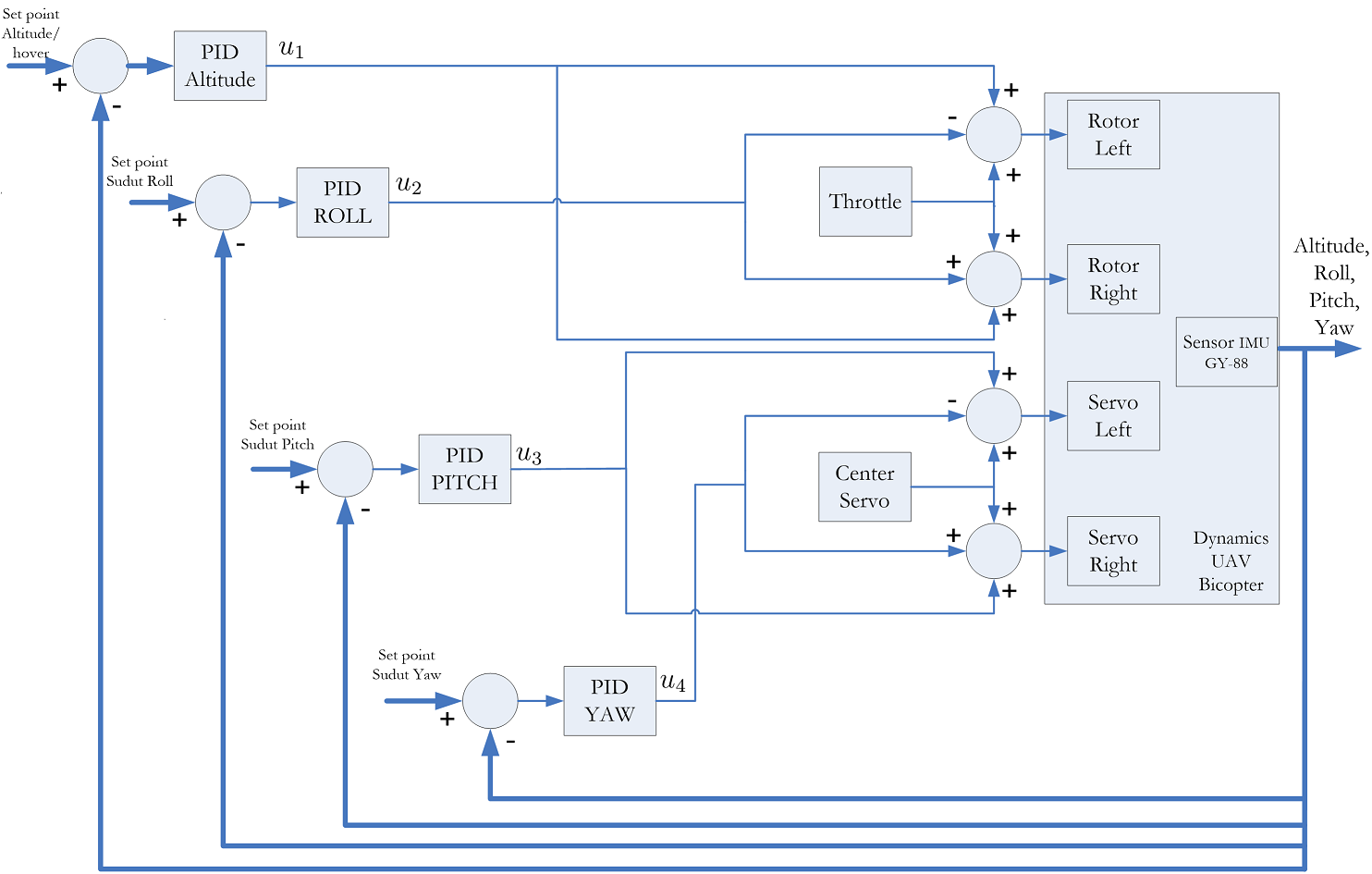}
	\caption{Control system block diagram for Bicopter flight stability.}
	\label{fig:gb2w25}
\end{figure*}

The block diagram of the closed-loop control system for the attitude stability of the Bicopter is presented in Fig. \ref{fig:gb2w25}. From this block diagram, it can be seen that there are four closed loops. The first is for the altitude control loop of the Bicopter, for the second, third and fourth loops are the attitude control of the Bicopter in the orientation of the motion angles of roll, pitch and yaw.

\subsection{PID Attitude Roll Control}

\begin{figure}[h]
	\begin{subfigure}{.5\textwidth}
		\centering
		\includegraphics[scale=0.2]{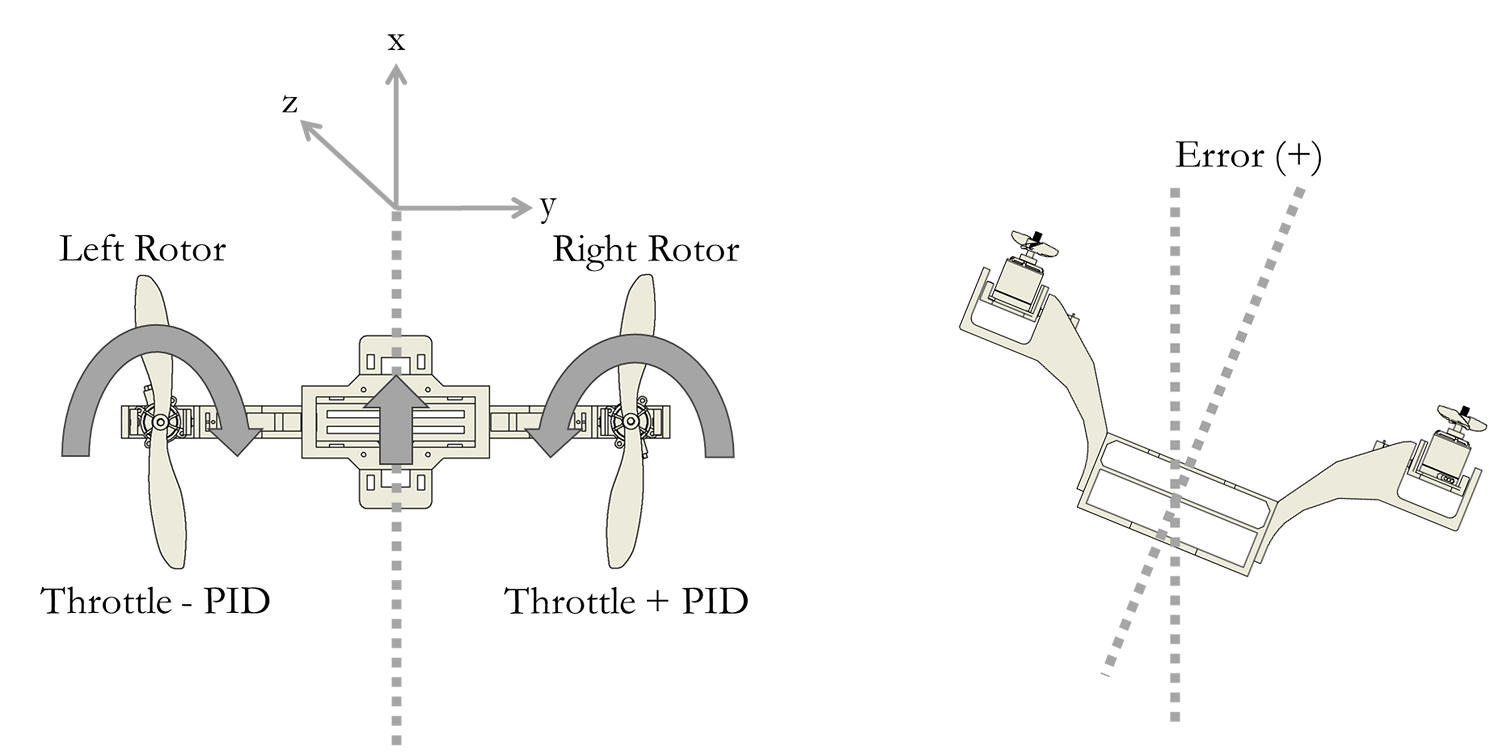}
		\caption{}
		\label{fig:BicopterRoll+}
	\end{subfigure}
	\begin{subfigure}{.5\textwidth}
		\centering
		\includegraphics[scale=0.2]{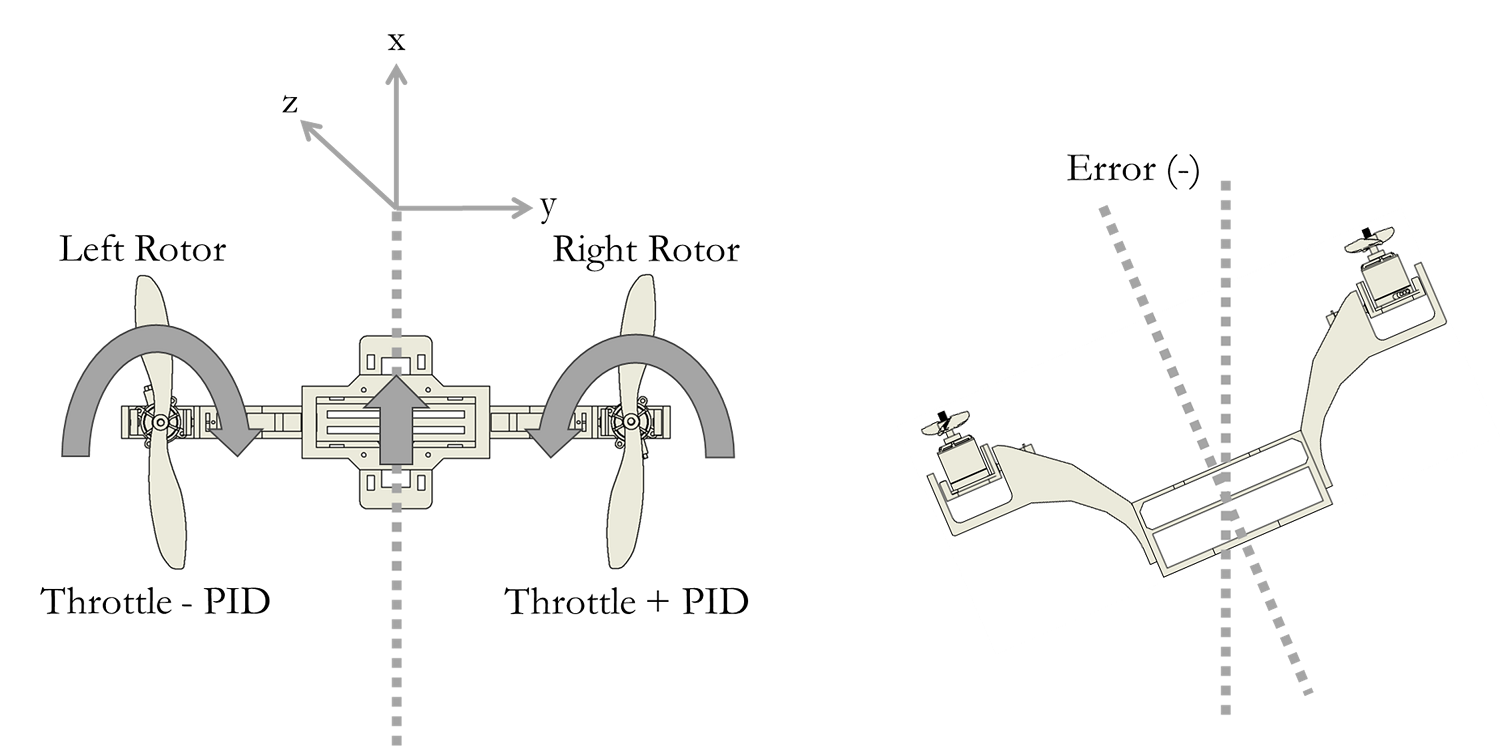}
		\caption{}
		\label{fig:BicopterRoll-}
	\end{subfigure}
	\caption{Bicopter attitude roll condition; (a) roll angle with an error ($+$) value produces a translational movement on the y-axis, which is shifted to the right, (b) roll angle with an error ($-$) value produces a translational movement on the y-axis, which is shifted to the left.}
	\label{fig:BicopterRoll}
\end{figure}

The attitude roll is a rotational movement of the Bicopter about the x-axis, which means this attitude movement will cause the translational displacement of the Bicopter on the y-axis to shift to the right and left. An illustration of the rolling motion of the Bicopter is shown in Fig. \ref{fig:BicopterRoll}. 

By providing an input reference, the set point signal (SP), in the form of a Bicopter roll angle of 0 degrees, then the deviation of the roll angle $\left ( \phi \right )$ from the reference roll angle $\left ( \phi _{r} \right )$ is defined as an error in Eq. (\ref{eror}). Furthermore, if you know the error value $\left ( e \right )$, then the differential error $\left (\frac{de(t)}{dt}  \right )$ can be calculated as shown in Eq. (\ref{deror}).

\begin{equation}
	e_\phi(t)=\phi -\phi _{r}
	\label{eror}
\end{equation}

\begin{equation}
	\frac{de_\phi(t)}{dt}=\dot{\phi} -\dot{\phi _{r}}
	\label{deror}
\end{equation}

Discrete PID control is another form of analog PID control (continues) in Eq. (\ref{pid_digital}) which is programmed and executed using a computer or microcontroller. The analog PID control must first be converted to a discrete form to implement discrete PID on a computer or microcontroller \cite{bhandari2022digital}. The formulation of discrete PID control can be seen in Eq. (\ref{pid_digital}) - (\ref{pid_diskrit}).

\begin{equation}
	u(t)=K_{p}e(t)+K_{i}\int_{0}^{t}e(t)dt+K_{d}\frac{de(t)}{dt} \\
	\label{pid_digital}
\end{equation}

With $K_{i}=\frac{1}{\tau_{i}}$ and $K_{d}=\tau_{d}$, the integral and differential forms can be written in discrete form as in Eq. (\ref{integral}) and Eq. (\ref{differensial}), so that they are obtained in the discrete PID control form in Eq. (\ref{pid_diskrit}). $e(k)$ is the current error, $e(k-1)$ is the previous error and $T$ is the sampling time.

\begin{equation}
	\int_{0}^{t}e(t)dt\approx T\sum_{0}^{k}e(k)
	\label{integral}
\end{equation}

\begin{equation}
	\frac{de(t)}{dt}\approx \frac{e{(k)}-e{(k-1)}}{T}
	\label{differensial}
\end{equation}

\begin{equation}
	\begin{split}
		u(k)=K_{p}e(k)+K_{i}T\sum_{0}^{k}e(k) \\
		+\frac{1}{T}K_{d}e{(k)}-e{(k-1)}
		\label{pid_diskrit}
	\end{split}
\end{equation}

In the case of controlling the attitude roll of the Bicopter using PID control, the output value of the PID roll controller in Eq. (\ref{pid_diskrit_roll}) will be added or subtracted from the given throttle value depending on the roll angle error condition, and this condition can be explained by looking at the illustration in Fig. \ref{fig:BicopterRoll}.

\begin{equation}
	\begin{split}
		u_\phi(k)=K_{p\phi} ~e_\phi(k)+K_{i\phi} ~T\sum_{0}^{k}e_\phi(k) \\
		+\frac{1}{T}K_{d\phi} ~e_\phi(k)-e_\phi(k-1)
		\label{pid_diskrit_roll}
	\end{split}
\end{equation}

\subsection{PID Attitude Pitch Control}

\begin{figure}[h]
	\begin{subfigure}{.5\textwidth}
		\centering
		\includegraphics[scale=0.2]{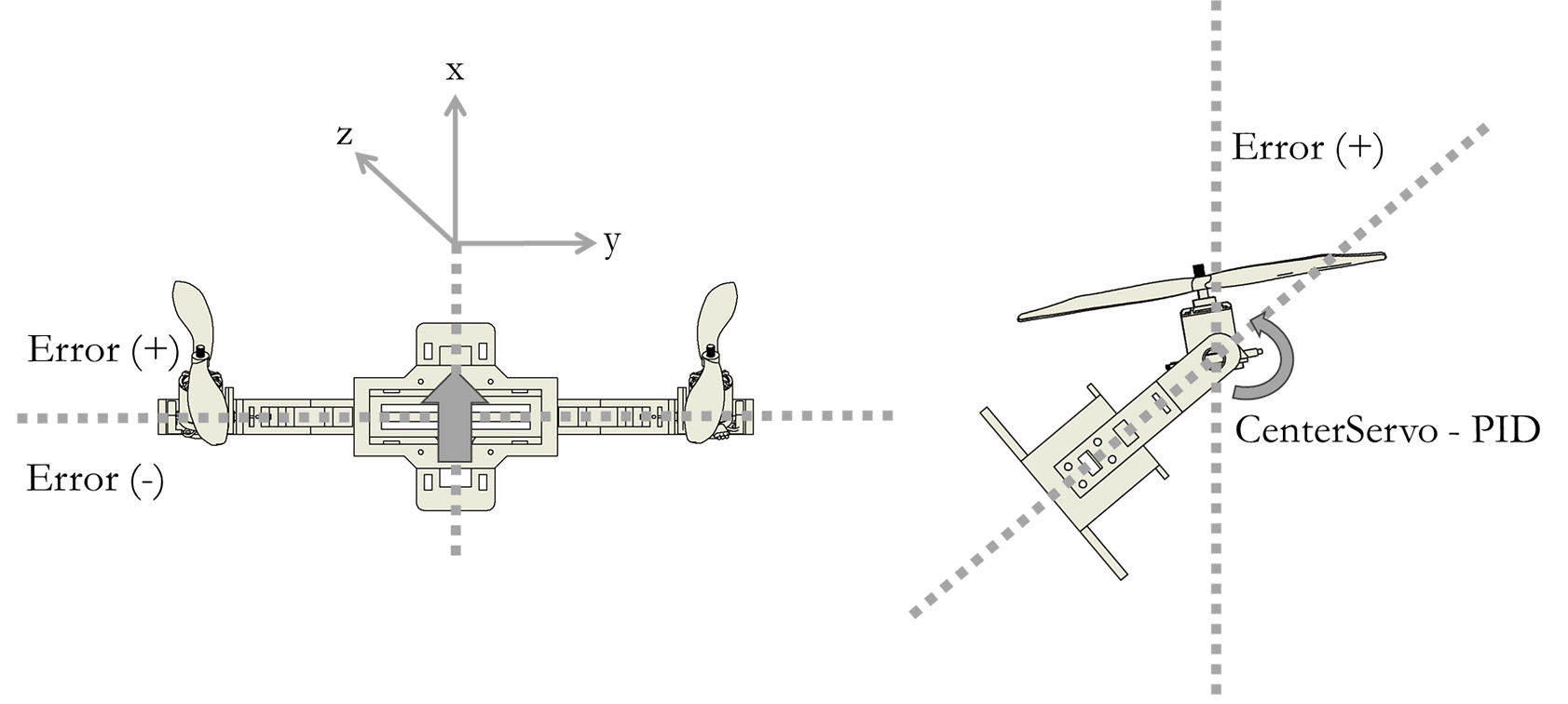}
		\caption{}
		\label{fig:BicopterPitch+}
	\end{subfigure}
	\begin{subfigure}{.5\textwidth}
		\centering
		\includegraphics[scale=0.2]{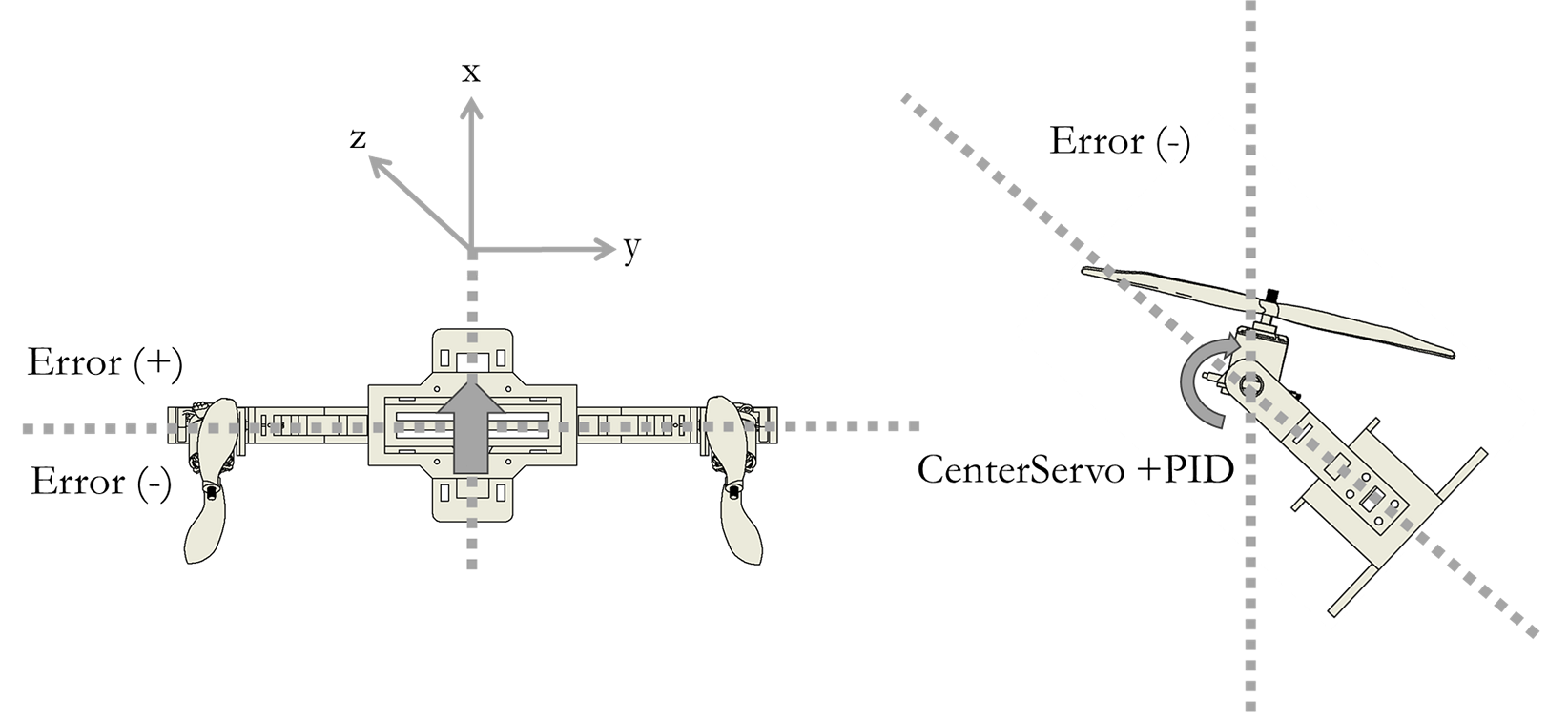}
		\caption{}
		\label{fig:BicopterPitch-}
	\end{subfigure}
	\caption{Bicopter attitude pitch condition; (a) pitch angle with error ($+$) value produces translational movement on the x-axis, which is shifted forward, (b) pitch angle with error ($-$) value produces translational movement on the x-axis, which is shifted backwards.}
	\label{fig:BicopterPitch}
\end{figure}

The attitude pitch is the rotational movement of the Bicopter about the \textit{y-axis}, which means this attitude will cause the translational displacement of the Bicopter on the \textit{x-axis} to shift forwards and/or backwards. In the case of controlling the attitude pitch of the Bicopter using PID control, the output value of the PID pitch controller in Eq. (\ref{pid_diskrit_pitch}) will be added or subtracted from the given CenterServo value depending on the pitch angle error condition; this condition can be explained by looking at the illustration in Fig. \ref{fig:BicopterPitch}.

\begin{equation}
	\begin{split}
		u_\theta(k)=K_{p\theta} ~e_\theta(k)+K_{i\theta} ~T\sum_{0}^{k}e_\theta(k) \\
		+\frac{1}{T}K_{d\theta} ~e_\theta(k)-e_\theta(k-1)
		\label{pid_diskrit_pitch}
	\end{split}
\end{equation}

\subsection{PID Attitude Yaw Control}

\begin{figure}[h]
	\begin{subfigure}{.5\textwidth}
		\centering
		\includegraphics[scale=0.29]{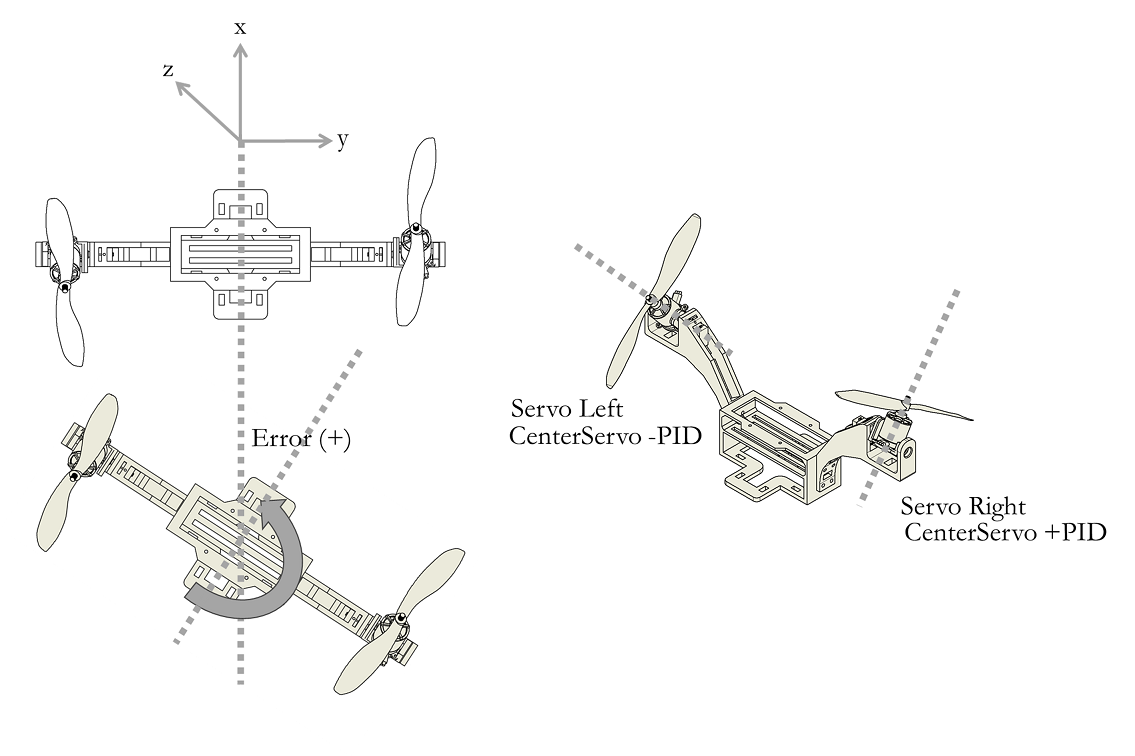}
		\caption{}
		\label{fig:BicopterYaw+}
	\end{subfigure}
	\begin{subfigure}{.5\textwidth}
		\centering
		\includegraphics[scale=0.29]{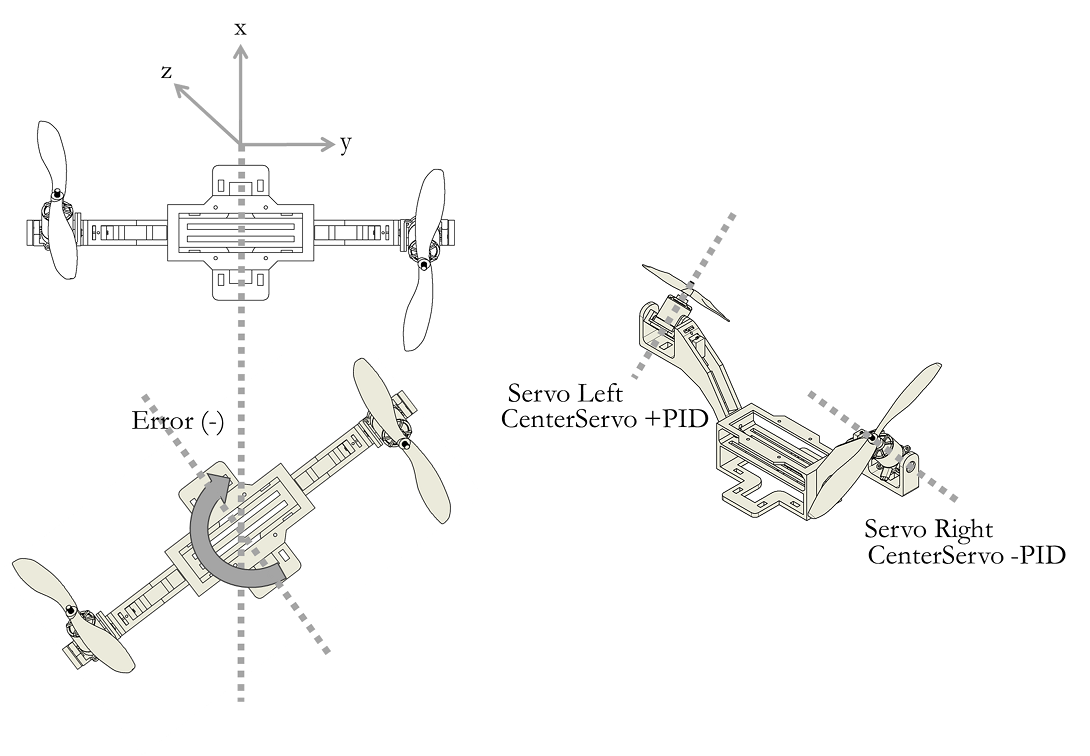}
		\caption{}
		\label{fig:BicopterYaw-}
	\end{subfigure}
	\caption{Bicopter attitude yaw condition; (a) yaw angle with error ($+$) value produces CCW rotational movement, (b) yaw angle with error ($-$) value produces CW rotational movement.}
	\label{fig:BicopterYaw}
\end{figure}

The attitude yaw is a rotational movement of the Bicopter about the z-axis, which means that this attitude movement will cause the rotational movement of the Bicopter to rotate clockwise (CW) or counterclockwise (CCW). In the case of controlling the attitude yaw Bicopter using PID control, the output value of the yaw PID controller in Eq. (\ref{pid_diskrit_yaw}) will be added or subtracted from the given CenterServo value depending on the yaw angle error condition, and this condition can be explained by looking at the illustration in Fig. \ref{fig:BicopterYaw}.

\begin{equation}
	\begin{split}
		u_\psi(k)=K_{p\psi} ~e_\psi(k)+K_{i\psi} ~T\sum_{0}^{k}e_\psi(k) \\
		+\frac{1}{T}K_{d\psi} ~e_\psi(k)-e_\psi(k-1)
		\label{pid_diskrit_yaw}
	\end{split}
\end{equation}

\subsection{PID Controller Tuning} \label{tuning kendali PID}

Before applying PID control parameters to control the attitude stability of the Bicopter, the tuning process of the PID attitude roll controller is carried out by simulating a dynamic model of the attitude roll of the Bicopter. Based on Eq. (\ref{eq3w43}), the dynamics of the Bicopter rolling motion can be obtained in the form of a double integrator. If the moment of inertia $I_{xx}$ is known to be $0.116\times10^{-3}$ and $L$ is $0.225$, the equation for the dynamic attitude roll transfer function can be obtained in Eq. (\ref{roll modelTF}). 

\begin{equation}
	\frac{\Phi}{U_2}=\frac{1939.7}{s^2}
	\label{roll modelTF}
\end{equation}

From Eq. (\ref{roll modelTF}) when changed in the model state space with suppose $\phi=y$, $u_2=u$, $x_1=y$, $x_2=\dot y$, $\dot{x}_1=\dot{y}$ dan $\dot{x}_2=\ddot{y}$ then obtained as follows:

\begin{equation}
	\begin{split}
		\dot{x}_{1}&=x_{2}\\
		\dot{x}_{2}&=1939.7 u\\
	\end{split}
	\label{xdotRoll}
\end{equation}

From Eq. (\ref{xdotRoll}) when arranged in the form $\dot{x}=Ax+Bu$, then obtained Matrix A and Matrix B as follows:

\begin{equation}
	\begin{split}
		\dot{x}=Ax+Bu\\
		A=\left [ \begin{matrix}
			0 & 1\\ 
			0 & 0
		\end{matrix} \right ]\left [ \begin{matrix}
			x_1\\ 
			x_2
		\end{matrix} \right ], B=\left [ \begin{matrix}
			0\\ 
			1939.7
		\end{matrix} \right ] u\\
	\end{split}
	\label{xdotRoll_state}
\end{equation}

It is known that the closed loop characteristic equation of the system in Eq. (\ref{xdotRoll_state}) can be obtained using the formula $det\left ( sI-\left ( A-BK \right ) \right )=0$, with the following description:

\begin{equation}
	\begin{split}
		det\left ( sI-(A-BK) \right )=0\\
		\left |s\left [ \begin{matrix}
			1   &0 \\ 
			0 & 1
		\end{matrix} \right ]  -\left ( \left [ \begin{matrix}
			0 &1 \\ 
			0& 0
		\end{matrix} \right ] - \left [ \begin{matrix}
			0\\ 
			1939,7
		\end{matrix} \right ]  \left [ \begin{matrix}
			K_1 & K_2
		\end{matrix} \right ]\right ) \right | =0\\
		\left | \left [ \begin{matrix}
			s & 0\\ 
			0 & s
		\end{matrix} \right ] - \left [ \begin{matrix}
			0 & 1\\ 
			0 & 0
		\end{matrix} \right ] -\left [ \begin{matrix}
			0 & 0 \\ 
			1939.7 K_1 & 1939.7 K_2
		\end{matrix} \right ]\right |=0\\
		\left | \left [ \begin{matrix}
			s & 0 \\ 
			0 & s
		\end{matrix} \right ] -\left [ \begin{matrix}
			0 & 1\\ 
			-1939.7K_1 & -1939.7K_2
		\end{matrix} \right ] \right |=0\\
		\left | \left [ \begin{matrix}
			s & -1\\ 
			1939,7K_1 & s+1939,7K_2
		\end{matrix} \right ] \right |=0\\
		s^{2}+\left ( 1939.7K_2 \right )s+1939.7K_1=0  	 
	\end{split}
	\label{xdotRoll_state_det}
\end{equation}

It is known that the characteristic equation of a closed two-loop system in general can be defined in Eq. (\ref{karakteristik}) where $\zeta$ is the damping ratio and $\omega_n$ is the natural frequency. By using the substitution method, it is possible to determine the gains of $K_1$ and $K_2$ according to $\zeta$ and $\omega_n$ based on the desired system performance.

\begin{equation}
	s^{2}+2\zeta \omega_n s+\omega_n^{2}=0
	\label{karakteristik}
\end{equation}

\begin{figure}[H]
	\centering
	\includegraphics[scale=0.2]{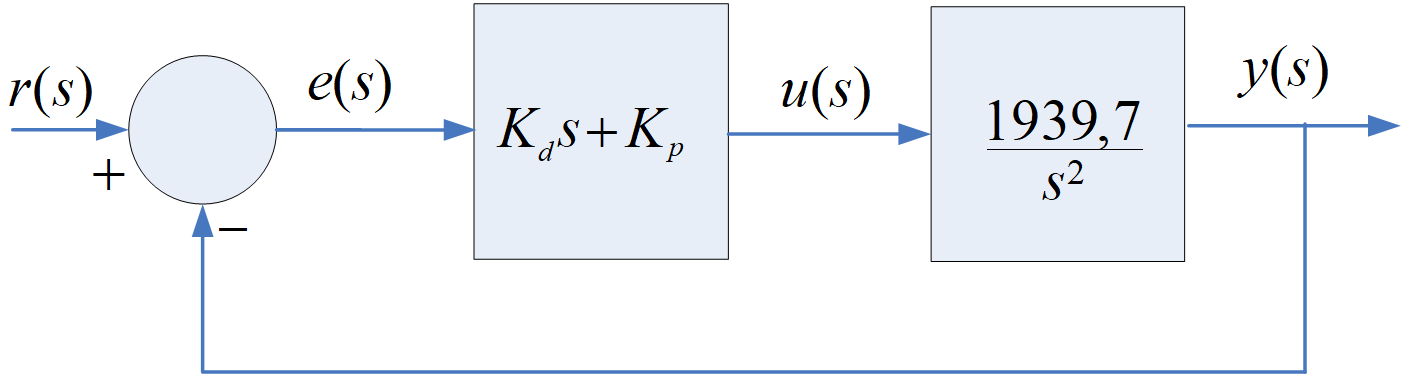}
	\caption{CLTF with PD controller.}
	\label{CLTFroll}
\end{figure}

The closed-loop transfer function (CLTF) dynamic attitude roll is planned in Eq. (\ref{roll modelTF}) using a proportional-differential (PD) controller, as shown in Fig. \ref{CLTFroll}. Equation (\ref{rumuscltfRoll}) explains the CLTF results.

\begin{equation}
	\begin{split}
		\frac{y(s)}{r(s)}&=\frac{\frac{1939,7K_ds+1939,7K_p}{s^{2}}}{1+\frac{1939,7K_ds+1939,7K_p}{s^{2}}} \\
		&=\underbrace{\frac{1939,7K_ds+1939,7K_p}{s^{2}+1939,7K_ds+1939,7K_p}}
		\label{rumuscltfRoll}
	\end{split}
\end{equation}

From Eq. (\ref{xdotRoll_state_det}) and Eq. (\ref{rumuscltfRoll}), it can be noticed that the value of $K_d = K_2$ and $K_p = K_1$, therefore if we want the system to have characteristics similar to Eq. (\ref{karakteristik}), then we will get $1939.7K_ds=2\zeta \omega_n s$ and $1939.7K_p=\omega_n^{2}$. If the planned closed loop system in Eq. (\ref{rumuscltfRoll}) has the characteristics of  $s^2+331s+1950=0$, $K_d$ and $K_p$ will be obtained in Eq. (\ref{KD}) and Eq. (\ref{KP}).

\begin{equation}
	\begin{split}
		K_d&=\frac{331s}{1939.7s}=0.1706\\
	\end{split}
	\label{KD}
\end{equation}

\begin{equation}
	\begin{split}
		K_p&=\frac{1950}{1939.7}=1.0053\\
	\end{split}
	\label{KP}
\end{equation}

\section{Results and Discussion}
\begin{figure*}
	\centering
	\includegraphics[scale=0.2]{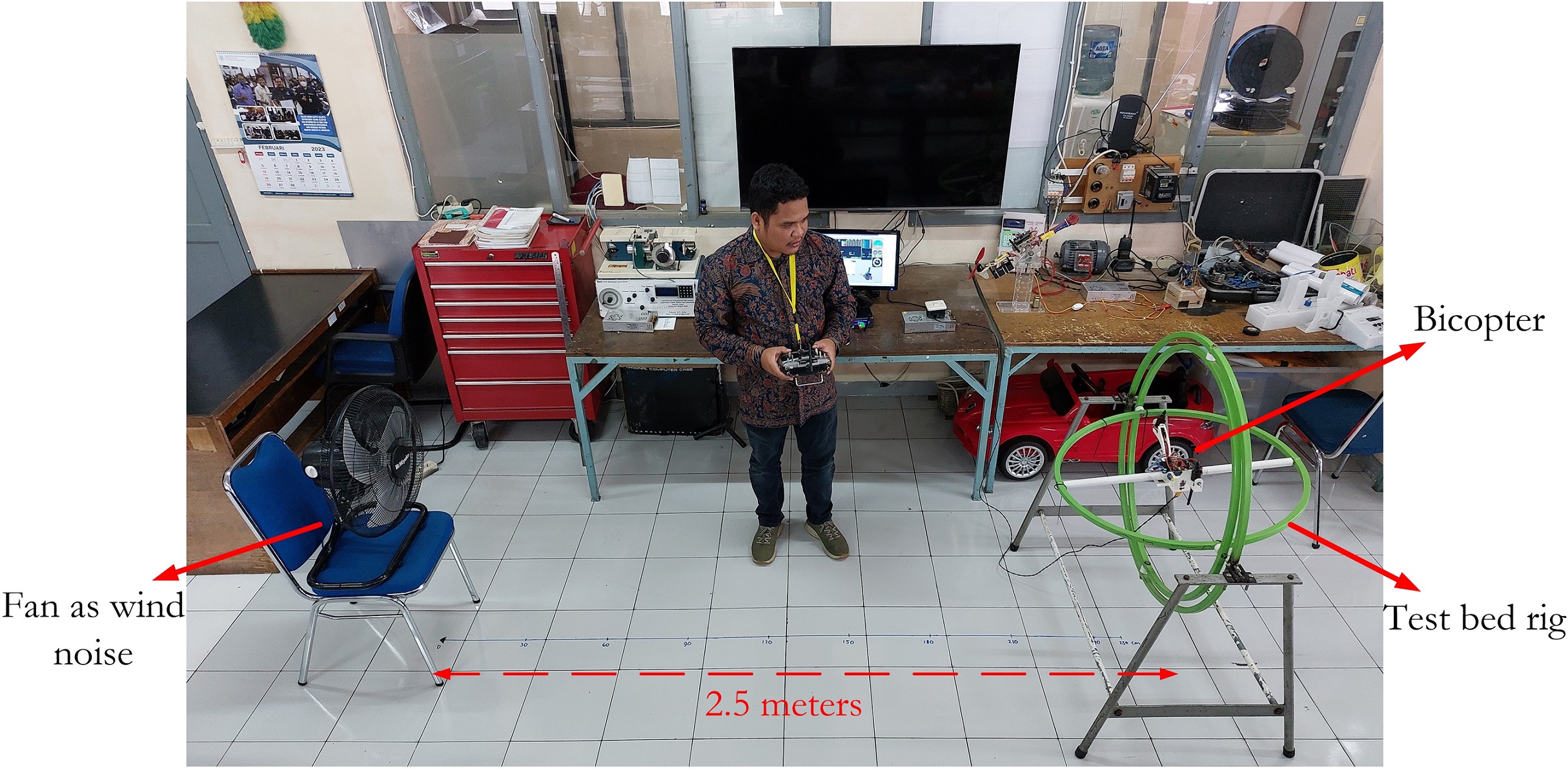}
	\caption{Bicopter test setup on a test bed rig with wind noise, the video recording of the test results can be seen in the following link: https://youtu.be/onJ8\_R9yVLQ .}
	\label{fig:setup pengujian}
\end{figure*}

A PID controller was implemented in the experiment to maintain a stable attitude from roll, pitch and yaw angles on the Bicopter. The test setup is carried out as shown in Fig. \ref{fig:setup pengujian}. With the help of the GUI, as presented in Fig. \ref{GUI}, the PID control parameter search process can be carried out using an experimental fine-tuning process by comparing the response results. This tuning process produces PID controller parameters described in Table \ref{tab:testbed}.

\begin{table}[H]
	\centering
	\caption{PID control parameters on attitude Bicopter using the test bed rig.}
	\label{tab:testbed}
	\begin{tabular}{llll}
		\toprule
		Parameter &  Roll &  Pitch &  Yaw \\ \midrule
		$K_p$        &        3.3       &     3.3           &       6.8       \\
		$K_i$        &       0.030        &       0.030          &    0.045        \\
		$K_d$        &         23      &          23      &           0   \\ \bottomrule
	\end{tabular}%
\end{table}

Furthermore, the PID control test is carried out by adding noise, by giving noise in the form of wind emitted using a fan. The wind given is regulated into three modes where the wind speed has been measured using a wind sensor in the form of an anemometer.

\begin{figure}[h]
	\centering
	\includegraphics[scale=0.55]{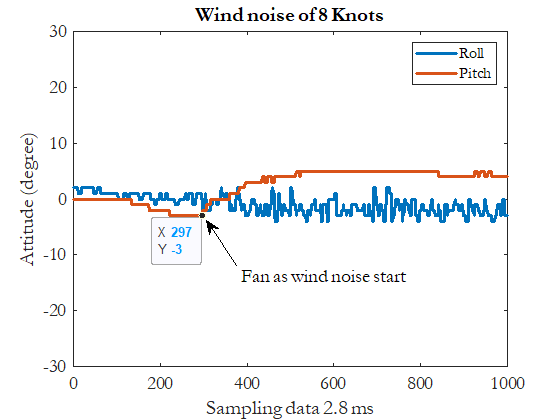}
	\caption{The attitude response of the Bicopter when given wind noise is around 8 Knots.}
	\label{fig:staticnoise8knots}
\end{figure}

The results of testing with noise were carried out in three modes, namely three types of wind strength. The first experiment, as shown in Fig. \ref{fig:staticnoise8knots} when the wind strength is set at a speed of about 8 Knots, it can be noticed that at about the 300th data or at the time (300 x 2.8 ms = 8.4 s) the attitude pitch angle of the Bicopter increases to $5^{\circ}$ means that the Bicopter experienced a condition of nose up of $5^{\circ}$. Up to 28 seconds of experiment, it can be seen that the attitude pitch condition of the Bicopter can be maintained. For the attitude roll condition, the Bicopter experiences shocks with a change in attitude roll of $\pm 3^{\circ}$.

\begin{figure}[h]
	\centering
	\includegraphics[scale=0.55]{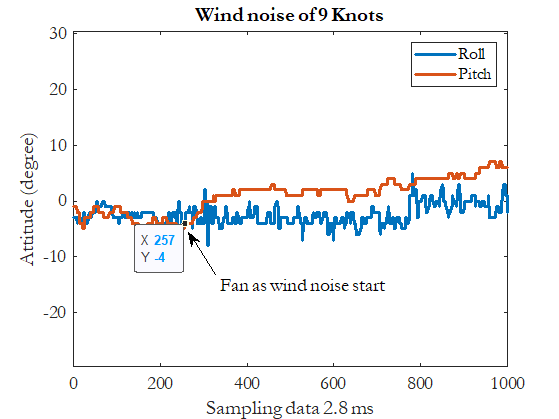}
	\caption{The attitude response of the Bicopter when given wind noise is around 9 Knots.}
	\label{fig:staticnoise9knots}
\end{figure}

In the second experiment by increasing the wind strength given by 9 Knots. In Fig. \ref{fig:staticnoise9knots} it can be seen that at around the 250th data or at the time (250 x 2.8 ms = 7.0 s) the attitude pitch angle of the Bicopter undergoes a nose up process until the 1000th data or at the time (1000 x 2.8 ms = 28.0 s) the attitude pitch angle of the Bicopter is around $7^{\circ}$ and the Bicopter experiences an attitude roll shock with a change of $\pm 4^{\circ}$.

\begin{figure}[h]
	\centering
	\includegraphics[scale=0.55]{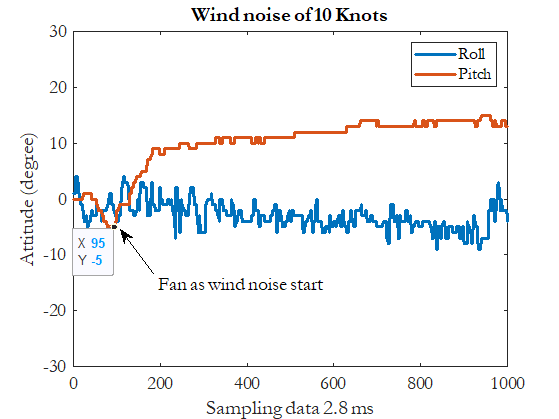}
	\caption{The attitude response of the Bicopter when given wind noise is around 10 Knots.}
	\label{fig:staticnoise10knots}
\end{figure}

In the third experiment, the wind strength was set at a speed of around 10 Knots, as seen in Fig. \ref{fig:staticnoise10knots}, the attitude pitch dipped to $11^{\circ}$ and also experienced an increase in attitude roll shocks with a change of $\pm 6^{\circ}$. The RMSE attitude roll and pitch values are presented in Table \ref{tab:RMSE static noise} when tested on a test bed. 

The Table \ref{tab:RMSE static noise} shows the RMSE attitude of a Bicopter when tested on a test bed with three variations of static noise. The RMSE attitude is a measure of how much the Bicopter's attitude (roll and pitch) deviates from the desired attitude. The higher the RMSE attitude, the more the Bicopter's attitude deviates from the desired attitude. The RMSE attitude of the Bicopter increases as the wind power from the fan increases. This is because the wind gusts cause the Bicopter to wobble, which increases the deviation of its attitude from the desired attitude. From  Table \ref{tab:RMSE static noise} also shows that the RMSE attitude of the Bicopter is higher for roll than for pitch. This is because the Bicopter is more susceptible to rolling than pitching and also because the Bicopter's rotors are located on the same plane, so they provide more lift for rolling than for pitching. Overall, the RMSE attitude of a Bicopter increases as the wind power from the fan increases and as the roll angle increases. This information can be used to design a Bicopter that is more stable in windy conditions.

\begin{table}[h]
	\centering
	\caption{RMSE attitude of Bicopter when tested on a test bed with three variations of static noise.}
	\label{tab:RMSE static noise}
	\begin{tabular}{llll}
		\toprule
		& \multicolumn{3}{c}{Wind power from the fan}     \\ \cline{2-4} 
		Attitude & 8 Knots & 9 Knots & 10 Knots \\ \midrule
		Roll     & 1.8868  & 2.7628  & 3.9183   \\
		Pitch    & 3.6764  & 4.2332  & 9.9868   \\ \bottomrule
	\end{tabular}%
\end{table}

\begin{figure}[h]
	\centering
	\includegraphics[scale=0.55]{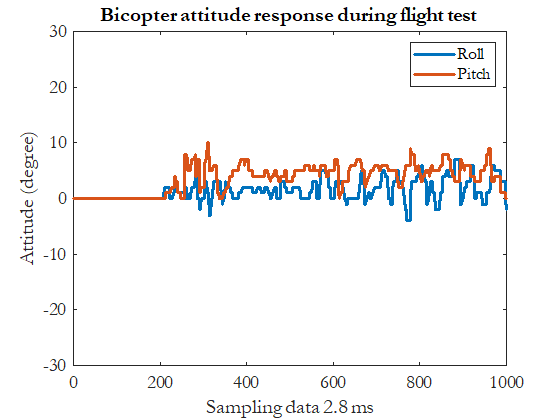}
	\caption{Attitude roll and pitch response during indoor Bicopter flight test.}
	\label{fig:terbangRollPitch}
\end{figure}

\begin{figure}[h]
	\centering
	\includegraphics[scale=0.55]{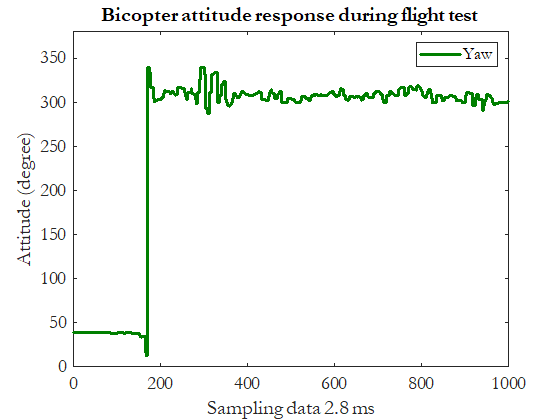}
	\caption{Attitude yaw response during indoor Bicopter flight test.}
	\label{fig:terbangYaw}
\end{figure}

\begin{figure}[h]
	\centering
	\includegraphics[scale=0.3]{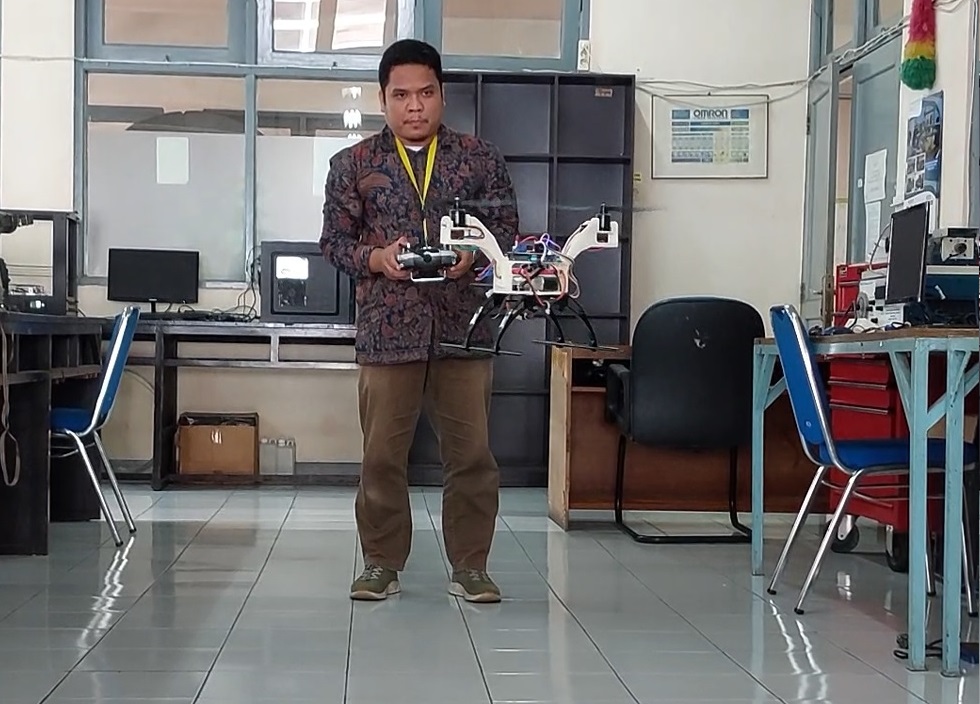}
	\caption{Bicopter indoor flight test, the video recording of the results of the Bicopter indoor flight test can be seen at the following link: https://youtu.be/rOh-Y5iN35g .}
	\label{fig:setup pengujian terbang}
\end{figure}

In subsequent tests, a PID controller was implemented to maintain stable roll, pitch and yaw angle attitudes on the Bicopter when flying indoors. The indoor flight test setup was carried out as shown in Fig. \ref{fig:setup pengujian terbang}. From the results of tuning the PID controller parameters using the experimental fine-tuning process, it was obtained parameters as presented in Table \ref{tab:pid terbang}. Figure \ref{fig:terbangRollPitch} shows the results of the attitude roll and pitch response during the flight test. Figure \ref{fig:terbangYaw} shows the results of the yaw attitude response. RMSE attitude roll and pitch values when tested in flight tests are presented in the Table \ref{tab:RMSE terbang}.

\begin{table}[h]
	\centering
	\caption{PID control parameters on attitude Bicopter flight test.}
	\label{tab:pid terbang}
	\begin{tabular}{llll}
		\toprule
		Parameter & Roll &  Pitch &  Yaw \\ \midrule
		$K_p$        &        1.3       &     1.3           &       0.1       \\
		$K_i$        &       0.030        &       0.108          &    0.010        \\
		$K_d$        &         20      &          12      &           16   \\ \bottomrule
	\end{tabular}%
\end{table}

The Figure \ref{fig:terbangRollPitch} shows the attitude response of a Bicopter during a flight test. The red line is the roll angle response, while the blue line is the pitch angle response. The x-axis shows the attitude data with time sampling 2.8 ms, and the y-axis shows the attitude angle in degrees. The oscillations in the roll and pitch angle responses are caused by the dynamics of the Bicopter and the disturbances that are present in the environment. In Fig. \ref{fig:terbangYaw}, the yaw angle response shows that the Bicopter oscillates between -35 degrees and 35 degrees.  The PID controller is designed to minimize these oscillations and keep the Bicopter's attitude close to the desired attitude. The sampling data of 2.8 ms indicates that the data was collected at a rate of 1000/2.8 = 357 Hz. The attitude response shows that the Bicopter is able to track the desired attitude with a reasonable degree of accuracy.

\begin{table}[h]
	\centering
	\caption{RMSE attitude of Bicopter during flight test.}
	\label{tab:RMSE terbang}
	\begin{tabular}{lc}
		\toprule
		Attitude & During flight test \\ \midrule
		Roll     & 2.3728                  \\
		Pitch    & 4.4219                  \\ \bottomrule
	\end{tabular}%
\end{table}

\section{Conclusion}

The results of implementing the PID controller on the attitude of the Bicopter have been tested for its durability using a test bed with various variations of wind strength. From the static noise test results, when the wind strength is given at a speed of 10 Knots, the RMSE value for attitude roll is 3.9183, and the RMSE value for attitude pitch is 9.9868. During the flight test, the PID controller maintained a stable attitude of the UAV Bicopter with an RMSE of 2.3728 for attitude roll and 4.4219 for attitude pitch. 

The mechanical design of the UAV Bicopter was developed using the concept of a “V” shaped frame with the aim that the center of mass (CoM) distribution of the UAV Bicopter can be in a position that causes the servomotor torque reaction to being parallel to the axis of rotation of the Bicopter when the attitude pitch angle moves. The electronic design of the UAV Bicopter was developed on the principle of low cost using the ATmega328P microcontroller.

\section*{Acknowledgment}
This work was supported by the Indonesian Postgraduate Domestic Education Scholarship (BPPDN) with contract number 2974/UN1.P.IV/KPT/DSDM/2019.

\bibliographystyle{IEEEtran}
\bibliography{ref}

\begin{thebibliography}{10}
\providecommand{\url}[1]{#1}
\csname url@samestyle\endcsname
\providecommand{\newblock}{\relax}
\providecommand{\bibinfo}[2]{#2}
\providecommand{\BIBentrySTDinterwordspacing}{\spaceskip=0pt\relax}
\providecommand{\BIBentryALTinterwordstretchfactor}{4}
\providecommand{\BIBentryALTinterwordspacing}{\spaceskip=\fontdimen2\font plus
\BIBentryALTinterwordstretchfactor\fontdimen3\font minus
  \fontdimen4\font\relax}
\providecommand{\BIBforeignlanguage}[2]{{%
\expandafter\ifx\csname l@#1\endcsname\relax
\typeout{** WARNING: IEEEtran.bst: No hyphenation pattern has been}%
\typeout{** loaded for the language `#1'. Using the pattern for}%
\typeout{** the default language instead.}%
\else
\language=\csname l@#1\endcsname
\fi
#2}}
\providecommand{\BIBdecl}{\relax}
\BIBdecl

\bibitem{kawasaki2015dual}
K.~Kawasaki, Y.~Motegi, M.~Zhao, K.~Okada, and M.~Inaba, ``Dual connected
  bi-copter with new wall trace locomotion feasibility that can fly at
  arbitrary tilt angle,'' in \emph{2015 IEEE/RSJ International Conference on
  Intelligent Robots and Systems (IROS)}.\hskip 1em plus 0.5em minus
  0.4em\relax IEEE, 2015, pp. 524--531.

\bibitem{saeed2018survey}
A.~S. Saeed, A.~B. Younes, C.~Cai, and G.~Cai, ``A survey of hybrid unmanned
  aerial vehicles,'' \emph{Progress in Aerospace Sciences}, vol.~98, pp.
  91--105, 2018.

\bibitem{ke2018design}
Y.~Ke, K.~Wang, and B.~M. Chen, ``Design and implementation of a hybrid uav
  with model-based flight capabilities,'' \emph{IEEE/ASME Transactions on
  Mechatronics}, vol.~23, no.~3, pp. 1114--1125, 2018.

\bibitem{qin2020gemini}
Y.~Qin, W.~Xu, A.~Lee, and F.~Zhang, ``Gemini: A compact yet efficient
  bi-copter uav for indoor applications,'' \emph{IEEE Robotics and Automation
  Letters}, vol.~5, no.~2, pp. 3213--3220, 2020.

\bibitem{albayrak2019design}
{\"O}.~B. Albayrak, Y.~Ersan, A.~S. Ba{\u{g}}ba{\c{s}}{\i}, A.~T.
  Ba{\c{s}}arano{\u{g}}lu, and K.~B. Ar{\i}kan, ``Design of a robotic
  bicopter,'' in \emph{2019 7th International Conference on Control,
  Mechatronics and Automation (ICCMA)}.\hskip 1em plus 0.5em minus 0.4em\relax
  IEEE, 2019, pp. 98--103.

\bibitem{zhang2016modeling}
Q.~Zhang, Z.~Liu, J.~Zhao, and S.~Zhang, ``Modeling and attitude control of
  bi-copter,'' in \emph{2016 IEEE International Conference on Aircraft Utility
  Systems (AUS)}.\hskip 1em plus 0.5em minus 0.4em\relax IEEE, 2016, pp.
  172--176.

\bibitem{hrevcko2015bicopter}
L.~Hre{\v{c}}ko, J.~Sla{\v{c}}ka, and M.~Hal{\'a}s, ``Bicopter stabilization
  based on imu sensors,'' in \emph{2015 20th International Conference on
  Process Control (PC)}.\hskip 1em plus 0.5em minus 0.4em\relax IEEE, 2015, pp.
  192--197.

\bibitem{panigrahi2021design}
S.~Panigrahi, Y.~S.~S. Krishna, and A.~Thondiyath, ``Design, analysis, and
  testing of a hybrid vtol tilt-rotor uav for increased endurance,''
  \emph{Sensors}, vol.~21, no.~18, p. 5987, 2021.

\bibitem{araar2014full}
O.~Araar and N.~Aouf, ``Full linear control of a quadrotor uav, lq vs h$\infty$,''
  in \emph{2014 UKACC International Conference on Control (CONTROL)}.\hskip 1em
  plus 0.5em minus 0.4em\relax IEEE, 2014, pp. 133--138.

\bibitem{cohen2020finite}
M.~R. Cohen, K.~Abdulrahim, and J.~R. Forbes, ``Finite-horizon lqr control of
  quadrotors on $ se\_2 (3) $,'' \emph{IEEE Robotics and Automation Letters},
  vol.~5, no.~4, pp. 5748--5755, 2020.

\bibitem{heng2015trajectory}
X.~Heng, D.~Cabecinhas, R.~Cunha, C.~Silvestre, and X.~Qingsong, ``A trajectory
  tracking lqr controller for a quadrotor: Design and experimental
  evaluation,'' in \emph{TENCON 2015-2015 IEEE region 10 conference}.\hskip 1em
  plus 0.5em minus 0.4em\relax IEEE, 2015, pp. 1--7.

\bibitem{okyere2019lqr}
E.~Okyere, A.~Bousbaine, G.~T. Poyi, A.~K. Joseph, and J.~M. Andrade, ``Lqr
  controller design for quad-rotor helicopters,'' \emph{The Journal of
  Engineering}, vol. 2019, no.~17, pp. 4003--4007, 2019.

\bibitem{saraf2020comparative}
P.~Saraf, M.~Gupta, and A.~M. Parimi, ``A comparative study between a classical
  and optimal controller for a quadrotor,'' in \emph{2020 IEEE 17th India
  Council International Conference (INDICON)}.\hskip 1em plus 0.5em minus
  0.4em\relax IEEE, 2020, pp. 1--6.

\bibitem{qin2022gemini}
Y.~Qin, N.~Chen, Y.~Cai, W.~Xu, and F.~Zhang, ``Gemini ii: Design, modeling,
  and control of a compact yet efficient servoless bi-copter,'' \emph{IEEE/ASME
  Transactions on Mechatronics}, 2022.

\bibitem{fedorov2015using}
D.~Fedorov, A.~Y. Ivoilov, V.~Zhmud, and V.~Trubin, ``Using of measuring system
  mpu6050 for the determination of the angular velocities and linear
  accelerations,'' \emph{Automatics \& Software Enginery}, vol.~11, no.~1, pp.
  75--80, 2015.

\bibitem{berkane2017design}
S.~Berkane and A.~Tayebi, ``On the design of attitude complementary filters on
  so(3),'' \emph{IEEE Transactions on Automatic Control}, vol.~63, no.~3, pp.
  880--887, 2017.

\bibitem{ngo2017experimental}
H.-Q.-T. Ngo, T.-P. Nguyen, V.-N.-S. Huynh, T.-S. Le, and C.-T. Nguyen,
  ``Experimental comparison of complementary filter and kalman filter design
  for low-cost sensor in quadcopter,'' in \emph{2017 International Conference
  on System Science and Engineering (ICSSE)}.\hskip 1em plus 0.5em minus
  0.4em\relax IEEE, 2017, pp. 488--493.

\bibitem{marantos2015uav}
P.~Marantos, Y.~Koveos, and K.~J. Kyriakopoulos, ``Uav state estimation using
  adaptive complementary filters,'' \emph{IEEE Transactions on Control Systems
  Technology}, vol.~24, no.~4, pp. 1214--1226, 2015.

\bibitem{fourati2012complementary}
H.~Fourati, N.~Manamanni, L.~Afilal, and Y.~Handrich, ``Complementary observer
  for body segments motion capturing by inertial and magnetic sensors,''
  \emph{IEEE/ASME transactions on Mechatronics}, vol.~19, no.~1, pp. 149--157,
  2012.

\bibitem{hemingway2018perspectives}
E.~G. Hemingway and O.~M. O’Reilly, ``Perspectives on euler angle
  singularities, gimbal lock, and the orthogonality of applied forces and
  applied moments,'' \emph{Multibody system dynamics}, vol.~44, pp. 31--56,
  2018.

\bibitem{brigante2011towards}
C.~M. Brigante, N.~Abbate, A.~Basile, A.~C. Faulisi, and S.~Sessa, ``Towards
  miniaturization of a mems-based wearable motion capture system,'' \emph{IEEE
  Transactions on industrial electronics}, vol.~58, no.~8, pp. 3234--3241,
  2011.

\bibitem{hua2013implementation}
M.-D. Hua, G.~Ducard, T.~Hamel, R.~Mahony, and K.~Rudin, ``Implementation of a
  nonlinear attitude estimator for aerial robotic vehicles,'' \emph{IEEE
  Transactions on Control Systems Technology}, vol.~22, no.~1, pp. 201--213,
  2013.

\bibitem{oh2014attitude}
H.-M. Oh and M.-Y. Kim, ``Attitude tracking using an integrated inertial and
  optical navigation system for hand-held surgical instruments,'' in \emph{2014
  14th International Conference on Control, Automation and Systems (ICCAS
  2014)}.\hskip 1em plus 0.5em minus 0.4em\relax IEEE, 2014, pp. 290--293.

\bibitem{katsuki2015rotation}
S.~Katsuki and N.~Sebe, ``Rotation matrix optimization with quaternion,'' in
  \emph{2015 10th Asian Control Conference (ASCC)}.\hskip 1em plus 0.5em minus
  0.4em\relax IEEE, 2015, pp. 1--6.

\bibitem{parwana2017quaternions}
H.~Parwana and M.~Kothari, ``Quaternions and attitude representation,''
  \emph{arXiv preprint arXiv:1708.08680}, 2017.

\bibitem{abedini2021robust}
A.~Abedini, A.~A. Bataleblu, and J.~Roshanian, ``Robust backstepping control of
  position and attitude for a bi-copter drone,'' in \emph{2021 9th RSI
  International Conference on Robotics and Mechatronics (ICRoM)}.\hskip 1em
  plus 0.5em minus 0.4em\relax IEEE, 2021, pp. 425--432.

\bibitem{bhandari2022digital}
P.~Bhandari and P.~Z. Csurcsia, ``Digital implementation of the pid
  controller,'' \emph{Software Impacts}, vol.~13, p. 100306, 2022.

\end{thebibliography}

\begin{IEEEbiography}[{\includegraphics[width=1in,height=1.25in,clip,keepaspectratio]{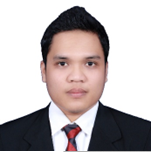}}]{Fahmizal} 
	received the B. Eng. degree in electrical engineering from Institut Teknologi Sepuluh Nopember (ITS), Indonesia in 2012. He graduated from Master of Science in Taiwan, precisely at National Taiwan University of Science and Technology (Taiwan Tech) in 2014. Start from 2015, he is a junior lecturer in the Department of Electrical and Informatics Engineering at Vocational College Universitas Gadjah Mada, Indonesia, in the specialist field of the control system and robotics.  Currently,  he is working toward the doctoral degree in  Department  of  Electrical  and  Information  Engineering,  Engineering  Faculty, Universitas  Gadjah  Mada,  Yogyakarta,  Indonesia. E-mail: fahmizal@ugm.ac.id
\end{IEEEbiography}

\begin{IEEEbiography}[{\includegraphics[width=1in,height=1.25in,clip,keepaspectratio]{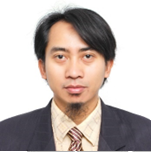}}]{Hanung Adi Nugroho} 
	received the B. Eng. degree in electrical engineering from Universitas Gadjah Mada, Indonesia, 2001 and Master of Engineering degree in Biomedical Engineering from the University of Queensland, Australia in 2005. In 2012, he received his Ph.D. degree in Electrical and Electronic Engineering from Universiti Teknologi Petronas, Malaysia. Currently, he is a Professor and also a Head of Department of Electrical Engineering and Information Technology, Faculty of Engineering, Universitas Gadjah Mada, Indonesia. His current research interests include biomedical signal and image processing and analysis, computer vision, medical instrumentation and pattern recognition. E-mail: adinugroho@ugm.ac.id 
\end{IEEEbiography}

\begin{IEEEbiography}[{\includegraphics[width=1in,height=1.25in,clip,keepaspectratio]{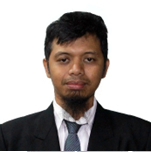}}]{Adha Imam Cahyadi} 
	received the B. Eng. degree in electrical engineering from University of Gadjah Mada, Indonesia in 2002. Then he worked as an engineerin industry, such as in Matsushita Kotobuki Electronics and Halliburton Energy Services for a year. He received the M. Eng. degree in control engineering from King Mongkuts Institute of Technology Ladkrabang, Thailand (KMITL) in 2005, and received the D. Eng. degree in control engineering from Tokai University, Japan in 2008. Currently, he is an Associate Professor at Department of Electrical Engineering and Information Technology, Faculty of Engineering, Universitas Gadjah Mada, Indonesia and a visiting lecturer at the Centre for Artificial Intelligence and Robotics (CAIRO), University of Teknologi Malaysia, Malaysia. His research interests include teleoperation systems and robust control for delayed systems especially process plant. E-mail: adha.imam@ugm.ac.id
\end{IEEEbiography}

\begin{IEEEbiography}[{\includegraphics[width=1in,height=1.25in,clip,keepaspectratio]{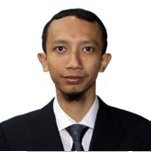}}]{Igi Ardiyanto} 
	received the B. Eng. degree in electrical engineering from University of Gadjah Mada, Indonesia in 2009. The M.Eng. and D. Eng. degrees in computer science and engineering from Toyohashi University of Technology (TUT), Japan in 2012 and 2015, respectively. He joined the TUT-NEDO (New Energy and Industrial Technology Development Organization, Japan) research collaboration on service robots, in 2011. He is now an Associate Professor at Department of Electrical Engineering and Information Technology, Faculty of Engineering, Universitas Gadjah Mada, Indonesia. He received several awards, including Finalist of the Best Service Robotics Paper Award at the 2013 IEEE International Conference on Robotics and Automation (ICRA 2013) and Panasonic Award for the 2012 RT-Middleware Contest. His research interests include planning and control system for mobile robotics, deep learning, and computer vision. E-mail: igi@ugm.ac.id
\end{IEEEbiography}

\EOD

\end{document}